\documentclass[prb,twocolumn,10pt,aps,showpacs,superscriptaddress]{revtex4-1}
\usepackage{graphicx}
\usepackage{amsmath}
\usepackage{amssymb}
\usepackage{bm}
\usepackage{dcolumn}
\usepackage{soul} 
\usepackage{psfrag}
\usepackage{float}
\usepackage{subeqnarray}

\usepackage{epstopdf}

\usepackage{url}
\usepackage[colorlinks=true,linkcolor=blue,citecolor=blue,urlcolor=blue,breaklinks]{hyperref}

\usepackage{placeins}

\begin{document}
	\nocite{apsrev41control}
\author{L.-F. Zhang}
\affiliation{Departement Fysica, Universiteit Antwerpen, Groenenborgerlaan 171, B-2020 Antwerpen, Belgium}
\author{L. Flammia}
\affiliation{Departement Fysica, Universiteit Antwerpen, Groenenborgerlaan 171, B-2020 Antwerpen, Belgium}
\affiliation{School of Science and Technology, Physics Division,
University of Camerino, 62032 Camerino, Italy}
\author{L. Covaci}
\affiliation{Departement Fysica, Universiteit Antwerpen, Groenenborgerlaan 171, B-2020 Antwerpen, Belgium}
\author{A. Perali}
\affiliation{School of Pharmacy, Physics Unit, University of Camerino,
62032 Camerino, Italy}
\author{M. V. Milo\v{s}evi\'{c}}\email{milorad.milosevic@uantwerpen.be}
\affiliation{Departement Fysica, Universiteit Antwerpen, Groenenborgerlaan 171, B-2020 Antwerpen, Belgium}

\title{Multifaceted impact of a surface step on superconductivity in atomically thin films}

\begin{abstract}
Recent experiments show that an atomic step on the surface of atomically thin metallic films can strongly affect electronic transport. Here we reveal multiple and versatile effects that such a surface step can have on superconductivity in ultrathin films. By solving the Bogoliubov-de Gennes equations self-consistently in this regime, where quantum confinement dominates the emergent physics, we show that the electronic structure is profoundly modified on the two sides of the step, as is the spatial distribution of the superconducting order parameter and its dependence on temperature and electronic gating. Furthermore, the surface step changes non-trivially the transport properties both in the proximity-induced superconducting pair correlations and the Josephson effect, depending on the step height. These results offer a new route to tailor superconducting circuits and design atomically thin hetero-junctions made of one same material.
\end{abstract}


\maketitle

\section{Introduction}

Over the last decade, atomically thin films were found to exhibit rich superconducting phenomena, often not achievable in their bulk counterparts \cite{wang_interface_2016, uchihashi_two-dimensional_2017, xu_large-area_2015, saito_highly_2016, brun_review_2017}.  This field of research was opened by discovery that Pb, In and Ga films can retain superconductivity down to thickness of few atomic monolayers \cite{guo_superconductivity_2004, ozer_hard_2006, zhang_superconductivity_2010, stepniak_atomic_2014, zhang_detection_2015}, in spite of expected detrimental effects of thermal and/or quantum fluctuations \cite{goldman_superconductorinsulator_2008, dubi_nature_2007}.  More recent discoveries include a strong enhancement of critical temperature $T_c$ in the one-unit cell thick FeSe films on SrTiO$_3$, above $100~\mathrm{K}$ compared to $8~\mathrm{K}$ of the bulk FeSe \cite{ge_superconductivity_2015};  the monolayers of NbSe$_2$ exhibit spin-momentum locking effect leading to a very high in-plane critical magnetic field \cite{xi_ising_2016};  the monolayer Tl-Pb compound hosts giant Rashba spin-split states, potentially useful for superconducting spintronics.  In all such crystalline and atomically thin materials superconductivity is known to be strongly affected by quantum confinement, leading to observable thickness-dependent quantum size effects \cite{guo_superconductivity_2004, shanenko_size-dependent_2006, croitoru_dependence_2007, romero-bermudez_shape_2014, romero-bermudez_size_2014,perali_molecule-like_2012,bianconi_shape_resonance_2014, innocenti_resonant_2010, doria_multigap_2016} and distinctly different electronic properties.  Understanding and controlling these is the key to engineering electronic devices with novel functionalities.

Recently, a step on the surface of atomically thin films, even if just one atom high, was found to strongly influence the electronic transport. Such surface steps not only change the overall electronic structure of the film, but also affect the range of proximity-induced superconducting correlations and the interplay of superconducting currents.  Refs. \onlinecite{brun_remarkable_2014, yoshizawa_imaging_2014} have demonstrated that an atomic surface step disrupts superconductivity, blocks supercurrents, pins Josephson vortices, and works as an intrinsic Josephson junction.  Furthermore, the surface step was recently found to terminate the propagation of the proximity-induced superconducting pair correlations \cite{kim_electrical_2016}.  Therefore, engineering the atomic steps on the surface of crystalline films is a definite new route to optimize and manipulate the superconducting properties or device performance at and below nanoscale.

In this paper, we show that above-mentioned effects are only some particular examples, as a surface step actually exhibits a multifaceted influence on the electronic, superconducting and transport properties of atomically thin films.  Due to the interplay between the quantum confinement effects and the scattering induced by the surface step, we find that physical properties on two sides of the step can be qualitatively and quantitatively different for both normal and superconducting state of the film.  In addition, the transport is also affected by quantum resonances related to film thickness, with performance characteristics tunable by the height of the step on the surface.  Our findings not only improve the understanding of the role of surface steps in superconductivity but also facilitate the further design of atomic-scale superconducting quantum devices.

The paper is organized as follows. Our theoretical formalism is outlined in Sec. II. Sec. III comprises the results of our numerical simulations of both stationary and transport properties of the superconducting films with nanoengineered surface steps, and ideas on local manipulation and fabrication of versatile junctions using surface steps. Sec. IV is devoted to the discussion on possible refinements of the model, and where and to which extent those refinements are expected to affect the main results of our calculations. The summary and a brief outlook are given in Sec. V.

\section{Theoretical formalism}\label{sec:2}

We employ the Bogoliubov-de Gennes (BdG) formalism, proven effective to study the interplay between superconducting phase and the quantum confinement effect in nanoscale samples \cite{shanenko_size-dependent_2006, croitoru_dependence_2007, zhang_unconventional_2012, shanenko_new_2007}.  Disorder, with its known scattering effects \cite{zhang_position-dependent_2015, zhang_tomasch_2015}, was not considered in this study, based on the experimentally shown robustness of the quantum-size effects in ultrathin Pb films \cite{bao_quantum_2005, eom_persistent_2006}. The BdG equations are written as:
\begin{eqnarray}
\label{BdG 1}  \left[K_0-E_F\right] u_n(\vec{r})+\Delta(\vec{r})v_n(\vec{r})&=&\mathcal{E}_n u_n(\vec{r}), \\
\label{BdG 2}  \Delta(\vec{r})^\ast u_n(\vec{r})-\left[K_0^\ast-E_F\right]
v_n(\vec{r})&=&\mathcal{E}_n v_n(\vec{r}),
\end{eqnarray}
where $K_0=-(\hbar\nabla)^2/2m+U(\vec{r})$ is the kinetic energy with $U$ being the one-particle potential and $E_F$ the Fermi energy, $u_n$($v_n$) are electron(hole)-like quasiparticle eigen-wavefunctions, and $E_n$ are the quasiparticle eigen-energies.  The pair potential $\Delta(\vec{r})$ is determined self-consistently from the eigen-wavefunctions and eigen-energies as:
\begin{equation}\label{OP}
\Delta(\vec{r})=g\sum\limits_{E_n<E_c}u_n(\vec{r}) v^\ast_n(\vec{r})[1-2f(E_n)],
\end{equation}
where $g$ is the coupling constant, $E_c$ the Debye cutoff energy, and $f(E_n)=[1+\exp(E_n/k_BT)]^{-1}$ the Fermi distribution function at temperature $T$.

We consider a laterally extended film with a surface step, schematically depicted in Fig.~\ref{Fig1}(a).  The thicknesses of the left and right side of the step are labelled $D$ and $d$, respectively.  We require that the wavefunctions $u_n(\vec{r})$ and $v_n(\vec{r})$ decay exponentially in the vacuum outside the film, by setting the potential to a large value, e.g.  $U(\vec{r}) = 20 E_F$.  Under this condition, we solve the BdG equations \eqref{BdG 1}-\eqref{OP} self-consistently by expanding $u_n$ and $v_n$ in terms of plane waves.  Then, periodic boundary conditions are automatically used in all ($x$, $y$ and $z$) directions.  It means that, in the $x$ direction, there will another step on both sides of the film.  In order to overcome the influence of the periodic images of the step, we consider a very long sample ($>100$ BCS coherence lengths $\xi_0$) so that the interaction between the periodically repeated steps is negligible. In the $y$ direction, the very high on-site potential $U(\vec{r})$ prevents any interaction with periodic images of the sample.

Since there exist by now a large variety of ultrathin superconducting structures, e.g. made of Pb, In, Ga, NbSe$_2$, we chose to keep the calculations generic and not include the specific band-structure of each material. Instead, we considered an isotropic quadratic dispersion and confirmed that the shown features remain robust for a wide range of parameters, i.e. $k_F \xi_0 = 2E_F/\pi \Delta_0 \approx 10-60$, where $k_F$ is the Fermi wave vector, $\xi_0$ and $\Delta_0$ are the coherence length and the bulk superconducting gap at $T=0$, respectively. Thus, without loss of generality, we set the parameters to: effective mass $m=1.5m_e$ ($m_e$ is the electron mass), $k_F \xi_0 \approx 30$ and $E_c/\Delta_0 \approx 10$.

\section{Results}\label{sec:3}

\subsection{Effect of the surface step on normal-state properties}
\begin{figure}
  \centering
  \includegraphics[width=\linewidth]{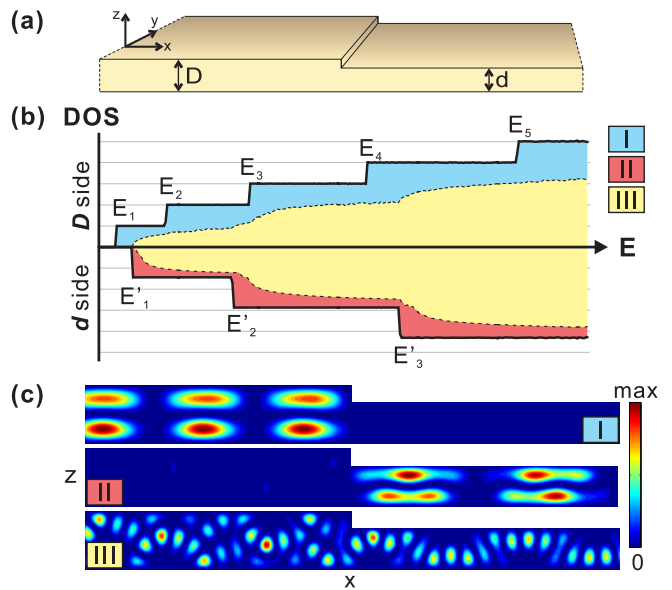}
  \caption{(Color online) (a) The oblique view of a film with surface step.  The thicknesses of two sides of the step are labelled as $D$ and $d$. The lateral dimensions of the film are taken large enough in the simulations and do not affect the presented results.  (b) Normal-state electronic structure for $D/d=10/7$.  The top panel shows the spatial average of LDOS on two sides of the step.  The color coding indicates the three classes of states (denoted $I$-$III$) and their portion in the total density. (c) The electronic probability density distributions for typical states $I$-$III$. }
  \label{Fig1}
\end{figure}

Since any effect of the surface step on electron wave functions may further manifest in the superconducting order parameter, we first examine the normal-state electronic structure, obtained by solving the single-electron Schr\"odinger equation, i.e. $K_0  \phi_l(\vec{r}) = E_l \phi_l(\vec{r})$.  We found that the normal-state electronic structure is well characterized by the spatial average of the LDOS (DOS) over $D$ and $d$ sides of the step.  As an exemplary case, Fig.~\ref{Fig1}(b) shows the DOS of both sides for $D/d=10/7$.  The film exhibits the standard 2D DOS of quantum wells on either side of the surface step, but with different characteristic energies, i.e. a staircase in DOS is observed at energies $E_j$ and $E'_{j'}$ on the corresponding side of the step.  Note that $E_j \rightarrow  \hbar^2(\pi j/D)^2/2m^*$ and $E'_{j'} \rightarrow  \hbar^2(\pi j'/d)^2/2m^*$ when the on-site potential $U \rightarrow \infty$ outside of the sample.  The calculated 2D DOS structure proves that our sample is sufficiently long so that the interaction between the periodically repeated steps is negligible, since no resonance peaks in DOS are observed.  To understand what causes the staircase behavior in DOS at energies $E_j$ and $E'_{j'}$, we examine also the normal state wavefunctions.  We find that the resulting normal states can be grouped into three classes [exemplified in Fig.~\ref{Fig1}(b)]: with probability density concentrated on thicker side (class I), on thinner side (II), or densities mixed across the step (III).  

The states of class I (II) are similar to the quantum-well states.  They emerge abruptly at threshold energies $E_j$ ($E'_{j'}$), and are responsible for the staircase rise of DOS [c.f. the color coding in DOS in Fig.~\ref{Fig1}(b)].  As a result, the electronic properties on one side of the step on the surface can be very different from those on the other, especially when the Fermi energy is close to $E_j$ or $E'_{j'}$.  At higher energy, the states of class I and II increasingly mix, and states of class III dominate the DOS.  Therefore the difference in DOS at the surface step becomes negligible for $j,j'\rightarrow \infty$, and homogeneous electronic properties in the film are recovered.  In other words, the surface step is particularly important for atomically thin films, and the peculiar electronic structure in that case is the consequence of the interplay between different quantum confinement effect on two sides of the step and the electron scattering at the surface step.  

\subsection{Superconducting properties and tunability of the system}

\begin{figure}
  \centering
  \includegraphics[width=\linewidth]{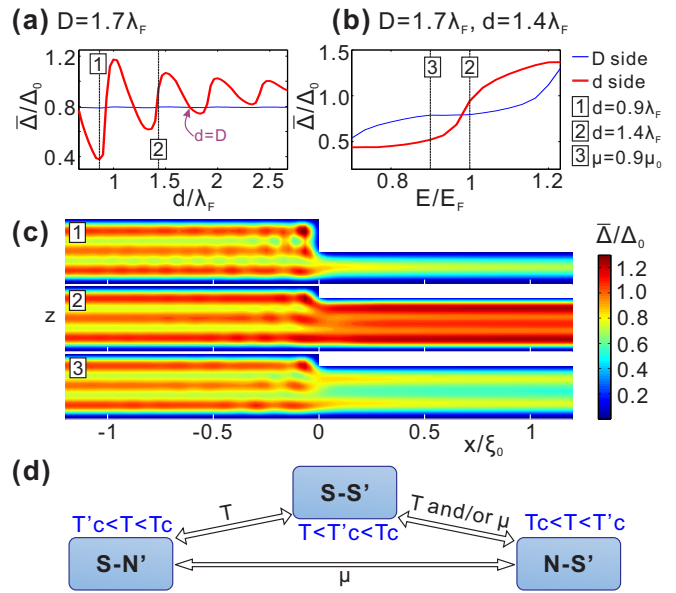}
  \caption{(Color online) Superconducting properties.  Panel (a) shows the spatial average of $|\Delta(x,z)|$ over two sides of the step, as a function of $d$ for $D=1.7\lambda_F$.  Panel (b) shows the same but as a function of $E$ for $D=1.7\lambda_F$ and $d=1.4\lambda_F$.  (c) The spatial profiles of the order parameter for the selected cases $1$-$3$.  (d) The chart for tuning the system into $S-S'$, $S-N'$ and $N-S'$ junctions by changing temperature $T$ and/or chemical potential $\mu$.}
  \label{Fig2}
\end{figure}

After understanding fundamental normal state properties of the system, we turn to the analysis of the superconducting state.  We first calculated the spatial average of the superconducting order parameter on two sides of the step, namely $\bar{\Delta}_D$ and $\bar{\Delta}_d$. Fig.~\ref{Fig2}(a) plots those quantities as a function of $d$ for fixed $D=1.7\lambda_F$, and reveals that $\bar{\Delta}_D$ does not change with $d$ while $\bar{\Delta}_d$ exhibits $d$-dependent oscillations due to quantum size effect. As a consequence, $|\Delta|$ on one side of the step on the surface can be very different from that on the other side, depending on the thicknesses $D$ and $d$ (relatively easily realized between 2 and 20 monolayers for e.g. Pb \cite{ozer_hard_2006}).  For example, for $D=1.7\lambda_F$, $|\Delta|$ in the $D$ side is higher than on the other side when $d=0.9\lambda_F$, while situation is reversed for $d=1.4\lambda_F$ [see cases $1$ and $2$ marked in Fig.~\ref{Fig2}(a)].  

The order parameter is not only sensitive to thickness $D$ and $d$, but can also be broadly tuned by the shift of Fermi energy $E_F$, via e.g. field effect ionic gating \cite{daghero_large_2012} (note however that the effect of ionic gating in reality could be more diverse than a mere change of $E_F$ \cite{jeong_suppression_2013, lu_full_2017}).  Fig.~\ref{Fig2}(b) shows the evolution of $\bar{\Delta}_D$ and $\bar{\Delta}_d$ with varied $E_F$, for $D=1.7\lambda_F$ and $d=1.4\lambda_F$.  As seen, $\bar{\Delta}_D$ and $\bar{\Delta}_d$ alternately dominate each other with changing $E_F$, as a consequence of normal state DOS being composed of staircase functions. Compared to the case $2$ at $E=E_F$ where $\bar{\Delta}_D < \bar{\Delta}_d$, in the case $3$ at $E=0.9E_F$ we realize $\bar{\Delta}_D > \bar{\Delta}_d$.  This feature is clearly seen in the contour plots of $|\Delta(x,y)|$ in Fig.~\ref{Fig2}(c) for the three selected cases. The noticeable Friedel-like oscillations on the scale of $\lambda_F$ in the vicinity of the step are due to the redistribution of charges, as discussed in Refs. \onlinecite{hayashi_low-lying_1998, dalla_torre_friedel_2016, machida_friedel_2003}.  Note that the spatial distribution of the order parameter, $|\Delta(x,y)|$, can be strongly affected by the microscopic inhomogeneities such as the atomic structure, disorder and other. However, the described effect of the surface step will remain present and distinct. In other words, the difference between the $|\Delta(x,y)|$ before and after the surface step would remain instructive, even though the details would change.

Employing the above features, one can realize S-S', S-N' and N-S' junctions (S,S' denote different superconductors, N and N' normal metals) in one same film with atomically defined step in thickness, nearly at will. In Fig.~\ref{Fig2}(d) we provide a schematic diagram for such tuning, done by changing chemical potential and/or temperature.  Due to different Andreev reflection and proximity effect in these three types of junctions, different electrical conductance can be realized, or used in thermal or electronic sensors. Additional functionalities of the device can be achieved in case of two (or more) steps on the film surface, in close proximity to each other.  Although similar tuning is feasible via proximity effect in SS’ bilayer structures \cite{de_gennes_boundary_1964}, our findings enable realization of multi-functional devices made of one same film, at the ultimate minimization of scale.

\begin{figure*}
  \centering
  \includegraphics[width=\textwidth]{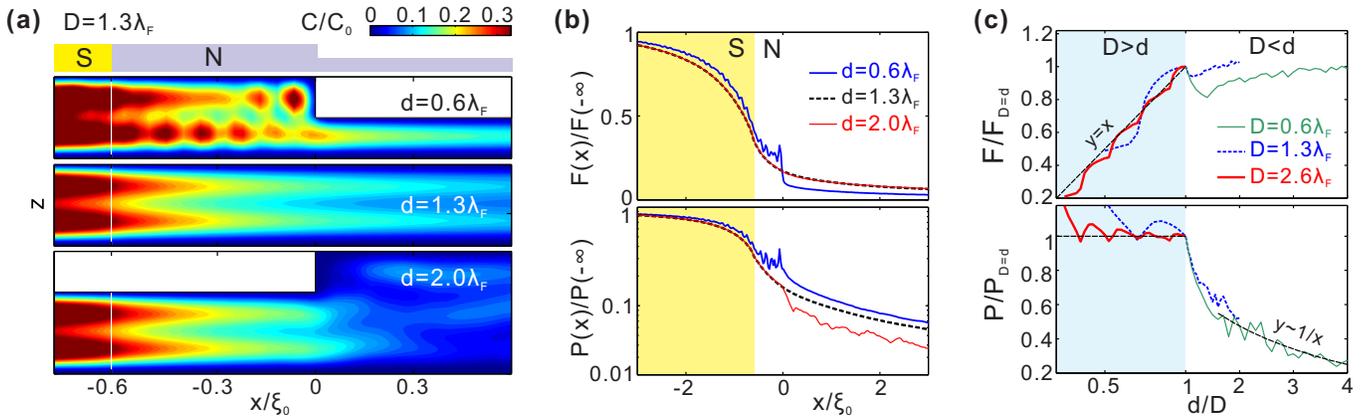}
  \caption{(Color online) The effect of a surface step on the proximity-induced superconducting pair correlations (PC).  (a) The spatial PC profile near the surface step for $d/\lambda_F=0.6$, $1.3$ and $2$, for fixed $D=1.3\lambda_F$.  $C_0$ is the bulk expectation value of the PC at $T=0$.  The S-N interface is located at $x=-0.6\xi_0$, as depicted in the top cartoon.  (b) $F(x)$, the total PC (integrated over $z$), and $P(x)$, the maximum of the PC over $z$, as a function of $x$ for the three cases shown in panel (a).  (c) $F/F_{D=d}$ and $P/P_{d=D}$ at $x=\xi_0$ as a function of $d/D$ for $D/\lambda_F=0.6$, $1.3$, and $2.6$.  The shaded area highlights the different behavior for $d<D$ compared to $d>D$ case.}
  \label{Fig3}
\end{figure*}

\subsection{Effect on transport properties - proximity and Josephson phenomena}

To reveal more facets of the influence of surface step on superconductivity in ultrathin films, we studied transport properties in the system. Motivated by the recent experiment of Ref. \onlinecite{kim_electrical_2016}, we study how the proximity-induced superconducting pair correlations change when crossing the step in thickness.  Since this is a system comprising superconducting and normal metal part, it is more convenient to express Eq.~\eqref{OP} as 
\begin{equation}\label{CP}
\Delta(\vec{r})=g(\vec{r})C(\vec{r})=g(\vec{r})\sum\limits_{E_n<E_c}u_n(\vec{r}) v^\ast_n(\vec{r})[1-2f(E_n)],
\end{equation}
where $C(\vec{r})$ are the superconducting pair correlations.  The coupling constant $g(\vec{r})$ is now spatially dependent, since it falls to zero in the normal metal part.  The results are presented in Fig.~\ref{Fig3}, for the case where the step itself is in the normal state, and superconducting correlations originate from distance $0.6\xi_0$ away from the step (i.e. $g(x>-0.6\xi_0)=0$). We find that superconducting transport from thicker to the thinner side of the film ($D>d$) strongly differs from the opposite case ($D<d$), due to different scattering induced by the surface step.  Fig.~\ref{Fig3}(a) shows comparatively the contourplots of order parameter near the step for cases $D>d$, $D=d$, and $D<d$. The normal state region is at $x>-0.6\xi_0$, and pair correlations is expected to decay with increasing $x$.  However, when $D>d$, the pair correlations are locally enhanced by reflection from the surface step, and exhibit pronounced oscillations in the film between the S-N interface and the surface step - notably different from the case $D=d$.  Surprisingly, the pair correlations beyond the surface step ($x>0$) are also enhanced compared to the $D=d$ case.  This is different from the observations in Ref. \onlinecite{kim_electrical_2016}, where the surface step terminated the propagation of the proximity-induced pair correlations.  On the other hand, for $D<d$, the pair correlations seem to decrease strongly beyond the surface step.  To visualize these features more clearly, we plot in Fig.~\ref{Fig3}(b) the pair correlations integrated over the thickness [$F(x)=\int \Delta(x,z) dz$] as well as the maximal local value of $\Delta(x,z)$ [$P(x)$] for each $x$, for the three cases considered in Fig.~\ref{Fig3}(a). We find that $F(x)$ is abruptly suppressed when crossing the surface step for $D>d$ whereas it does not change for $D<d$, compared to the case of $D=d$.  The behavior of $P(x)$ is opposite from $F(x)$ as seen also in Fig.~\ref{Fig3}(a), where $P(x)$ is enhanced when crossing the surface step for $D>d$ whereas it is suppressed for $D<d$, compared to the case of $D=d$. Note that oscillations in $F(x)$ and $P(x)$  are not due to the numerical accuracy but appear due to scattering at the interface and are pronounced in the thicker part of the sample.

In Fig.~\ref{Fig3}(c) we plot $F(x)$ and $P(x)$ at $x=\xi_0$ as a function of $d$, for different selected $D$ values, in order to devise the general trend with respect to the role of the surface step.  When $d<D$, the step blocks the propagation of the Cooper pairs and $F(x=\xi_0)$ increases linearly as $d\rightarrow D$. When $d>D$, the surface step does not block the propagation and $F(\xi_0)$ weakly increases with $d$ until saturation for $d \gg \xi_0$.  $F(\xi_0)$ still shows thickness-dependent oscillatory quantized behavior, especially for $d<D$.  Due to quantum confinement, the local density of the order parameter [$P(\xi_0)$] can be in resonant situation and thus enhanced for some $d$ when $d>D$.  When $d>D$, $P(x)$ on $d$ side is always proportional to $1/d$ so pair correlation density decreases fast when crossing the step on the film.

\begin{figure}
	\centering
	\includegraphics[width=\columnwidth]{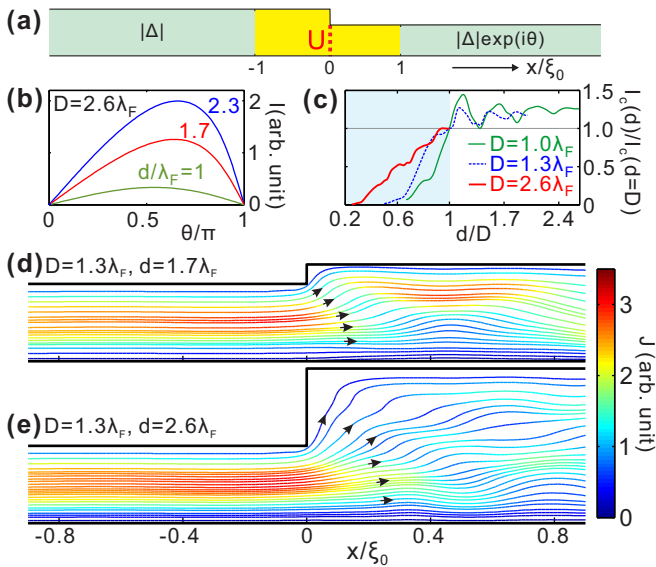}
	\caption{(Color online) (a) Schematic picture of a Josephson junction in an ultrathin film with a step, with additional potential barrier $U$.  (b) The current-phase relation for $D=2.6\lambda_F$ and $d/\lambda_F=1$, $1.7$ and $2.3$.  (c) Critical current $I_c$ as a function of $d/D$ (scaled to $I_c$ for $d=D$), for $D/\lambda_F=1$, $1.3$, and $2.6$.  The shading highlights the very different behavior observed for $d<D$ versus the case of $d>D$.  (d) and (e) show the supercurrent flow through the junction for $d/\lambda_F=1.7$ and $2.6$, respectively, for $D=1.3\lambda_F$.}
	\label{Fig4}
\end{figure}
Our results show that the effect of the surface step on the proximity-induced order parameter is more diverse than the bare termination observed by scanning tunneling microscopy in Ref. \onlinecite{kim_electrical_2016}.  The latter pertinent observation is suggestive of additional scattering at the experimentally realized surface step compared to our considerations.  We therefore introduce an additional potential barrier at the surface step [shown in Fig.~\ref{Fig4}(a)], because the surface step breaks the lattice translation symmetry, which can significantly modify the electronic structure and shift the chemical potential.  Such a potential barrier always has a detrimental effect on the proximity-induced pair correlations at the surface step. However, in that case we have additional new physics stemming from the fact that potential barrier at the step can be considered as a weak link.  To capture those effects, we study DC Josephson tunneling by imposing a phase difference $\theta$ between the two sides of the step on the surface \cite{covaci_proximity_2006}.  In practice, we set $\Delta(x<-\xi_0)=|\Delta|$ and $\Delta(x>\xi_0)=|\Delta|e^{i\theta}$ on the weak link of width $2\xi_0$ where additional potential barrier is applied [see Fig.~\ref{Fig4}(a)].  Then, the resulting supercurrent density is calculated as:
\begin{eqnarray}\label{current}
\vec{J}(\vec{r}) &= \frac{e\hbar}{2mi} \sum_{E_n<E_c}\left \{ f(E_n)u_n^*(\vec{r})\bigtriangledown u_n(\vec{r}) \right.\nonumber \\
&+ \left. [(1-f(E_n)]v_n(\vec{r})\bigtriangledown v_n^*(\vec{r})-h.c.\right \}.
\end{eqnarray}
The potential barrier $U'$ is chosen such that the critical Josephson current $I_c$ can always be found for $0<\theta<\pi$, as seen from the $I-\theta$ relation in Fig.~\ref{Fig4}(b).  In this work, the Gaussian potential barrier $U'$ is used, i.e. 
$U'(x)=U_0 \mathrm{exp}(-\frac{x^2}{2\sigma^2})$, where $U_0=2E_F$ and $\sigma=0.1\xi_0$.  Fig.~\ref{Fig4}(c) summarizes the effect of the surface step on the Josephson effect, where $I_c$ was plotted as a function of $d$ for different $D$.  Similarly to the proximity-induced order parameter, the effect on Josephson current is profoundly different for $d<D$ and $d>D$.  For $d<D$, $I_c$ increases nearly linearly with $d$.  For $d>D$, $I_c$ oscillates with $d$ due to quantum-size effect, until convergence.  These oscillations in $I_c$ indicate significant change in tunneling between two sides of the step.  This change is due to the mismatch of Fermi velocities, and more importantly in this case, due to details of the coupling between the sub-bands and the inter-band scattering.  Figs.~\ref{Fig4}(d) and (e) show the streamlines of the supercurrent inside the junction, for two thicknesses $d$, both larger than $D$.  Both considered cases exhibit very inhomogeneous flow of current, due to quantum confinement.  In Fig.~\ref{Fig4}(d), the main current channel shifts from the center/bottom of the film to the top when passing the step on surface.  In contrast, Fig.~\ref{Fig4}(e) shows the case where main current channel remains in the bottom half of the film before and beyond the step.  For that reason, the latter case exhibits facilitated tunneling, leading to higher critical current.  This is yet another example of a useful property in superconducting devices that can be tuned to very different regimes by atomically small steps in ultrathin films.  

\section{Discussion}\label{sec:4}

Arguably the main result of this paper is that superconductivity is strongly and diversely modified on two sides of the surface step, enabling realization of multi-functional superconducting hetero-junctions.  This fundamentally depends only on the manner in which superconductivity is thickness dependent - microscopic details do not change the generic picture, although some features may become weaker in cases of very small atomic steps and e.g. presence of considerable disorder. 

All our results are obtained by using a simplified model for superconducting thin films, where we considered only the quantum confinement effect on electrons, leading to the quantization of electron spectrum into different subbands \cite{shanenko_atomically_2015}.  We also assumed the pairing interaction to be the same as in the bulk material, defined by a local interaction with a constant pairing strength $V_0$, i.e. $V(r,r')=V_0 \delta(r-r')$.  Finally, we used a high surface potential barrier to confine the electronic motion in the film. Each of these assumptions can be improved so that the model becomes more realistic.  For example, electron-phonon pairing interaction in thin films is known to deviate from the bulk behavior,\cite{luh_large_2002, chen_superconducting_2012-1} as was experimentally reported for Pb atomic thin films.\cite{zhang_band_2005-1}  In what follows, we discuss how this and other refinements in the model affect the main results of our work.

First, the pairing interaction deviates from the bulk when electrons are confined in a film.  In BCS theory the pairing interaction Hamiltonian is $H_{int}=-\sum_{p,q} c^{\dagger}_{p\uparrow} c^{\dagger}_{p\downarrow} \hat{V}_{p,q} c_{q\downarrow} c_{q\uparrow}$ with $c^{\dagger}_{p\uparrow}$ and $c_{q\uparrow}$ the usual quasiparticle creation and annihilation operators.  The interaction matrix element is $\hat{V}_{p,q}=V_0 \int d^3\vec{r} |\Psi_p(\vec{r})|^2 |\Psi_q(\vec{r})|^2 $ with $\Psi_p(\vec{r})$ the eigen-wavefunctions.  Therefore the pairing interaction matrix elements depend not only on the interaction strength $V_0$ but also on the eigen-wavefunctions $\Psi_p(\vec{r})$.  In a film with thickness $L_z$, $\Psi_p(\vec{r}) \propto e^{i(k_xx+k_yy)} \mathrm{sin}(\pi p z/L_z)$ so that $\hat{V}_{p,q}=V_0/\mathcal{V}(1+\frac{1}{2}\delta_{pq})$ in the system with volume $\mathcal{V}$.  Using this model, Thompson and Blatt obtained the saw-tooth pattern in $T_c$ as thickness was varied.\cite{blatt_shape_1963}  Since their pairing interaction is stronger than ours, i.e. $\hat{V}_{p,q}=V_0/\mathcal{V}$, the $T_c$ is expected to be higher and the saw-tooth oscillations in $T_c$ are expected to be more pronounced.  As a result, the thickness dependence of superconductivity is more significant.  It is worth mentioning that Ref.~\onlinecite{valentinis_rise_2016, valentinis_bcs_2016} also take the energy dependence of the density of states into account, and find the saw-tooth pattern in $T_c$ with film thickness is suppressed.  However, that would not change our results because this effect has been included automatically in our BdG calculation.

Second, the pairing interaction also depends on the quantization of the phonon spectrum due to the quantum confinement effect.  Ref.~\onlinecite{hwang_role_2000} took this into account by setting a cutoff in the number of phonon modes.  Then, only the phonon modes $l=1,...,l_{max}$ contribute to the pairing interaction, and the $l_{max}$, the maximal allowed value of $l$, is proportional to the film thickness $d$ and the Debye cutoff energy $\hbar \omega_D$.  As a result, superconductivity is suppressed due to the less effective number of phonon modes.  However, the saw-tooth oscillations in $T_c$ are not affected.  In addition, resonance features appear more frequently with the variation of the film thickness.  Therefore, the thickness dependence of superconductivity remains significant, hence all our findings qualitatively hold.
  
Third, Ref.~\onlinecite{saniz_confinement_2013} considered the effect of quantum confinement on the electron-phonon coupling strength.  In contrast to the studies mentioned previously, the authors derived the phonon-mediated pairing interaction beyond the contact potential approximation in the frame of the Green's function approach.  They found the pairing interaction depends on two effects: 1) the screened Coulomb interaction; and 2) the number of phonon modes.  The former one drops remarkably when the number of occupied electron subbands increases by one, leading to the suppression in $T_c$.  On the other hand, the latter one rises when the number of phonon modes increases by one.  Thus those two effects compete with each other and lead to more pronounced $T_c$ oscillations with varying film thickness.  Therefore, the thickness dependence of superconductivity is still present and is pronounced, though details may be strongly changed.

Finally, a very high surface potential barrier is used in our model in order to create the quantum confinement effect on electrons.  For that reason, our results do not exhibit essentially different behavior from the results obtained if an infinite potential barrier is used.\cite{blatt_shape_1963, croitoru_dependence_2007}  However, a relatively low potential is likely a more realistic choice, being set by the work function of the material.  In films, the work function is usually suppressed compared to the bulk due to the finite size effect.\cite{qi_atomic-layer-resolved_2007}  Refs.~\onlinecite{yu_consistent_1976, romero-bermudez_size_2014, valentinis_bcs_2016} already studied superconductivity in thin films with a relatively low surface potential barrier, and found that the shape resonances of $\Delta$ and $T_c$ are still significant but the envelope curve exhibits a reduction with decreasing film thickness.  This reduction is induced by the leakage of the electronic wave function across the film surface.  Therefore, the thickness dependence of superconductivity will remain evident, though less significant.  

When all the aspects are taken into account, we render our main results qualitatively robust against above improvements in the model, albeit fine details may change and some effects can be weaker.

One of the other important results of our work is that the transport ability (e.g. in the propagation of the superconducting pair correlations or the supercurrent) across the surface step increases linearly as $d \rightarrow D$, with a quantized oscillatory background.  This feature depends mainly on the ratio between the thicknesses of two sides of the surface step and also depends on whether there is a momentum mismatch at the interface where the step is located.  As long as the step exists, both these conditions are satisfied.  Therefore, the above refinements of the model would not affect the reported transport behavior, characteristic of ultrathin films with a surface step. 

\section{Conclusion}\label{sec:5}

In summary, we showed that surface steps in atomically thin films can lead to a multitude of pronounced and diverse effects on superconductivity. Even if small, such steps strongly modify the effects of quantum confinement in ultrathin films, while also causing significant electronic scattering. We reveal how these effects are tuned by the thickness, temperature, and/or electronic gating, thus enabling the engineering of S-S', S-N' and N-S junctions at the step, nearly at will. The transport properties and the proximity effect across the step(s) also exhibit versatile behavior, even opposite regimes, depending on the thickness. In applied magnetic field, the discussed features will directly affect the properties of vortices (in the film \cite{zhang_superconductivity_2010}, near the S-S' step \cite{brun_remarkable_2014,  yoshizawa_imaging_2014}, or in an S-N-S' junction between the adjacent steps \cite{roditchev_direct_2015}), and their behavior in applied current. Further rich physics can be stimulated in these systems by tuning the spin-orbit coupling and applying an in-plane magnetic field \cite{xi_ising_2016}.

The needed crystalline thin films are readily experimentally available, and that for a range of different materials. Our findings thus present a first step towards ultrathin superconducting circuitry with functionality locally and broadly crafted by 3D atomistic engineering and quantum effects. 

\begin{acknowledgments}
This work was supported by the Research Foundation Flanders (FWO-Vlaanderen), the Special Research Funds of the University of Antwerp (TOPBOF project) and the Italian MIUR through the PRIN 2015 program (Contract No. 2015C5SEJJ001).
\end{acknowledgments}


\begin{thebibliography}{51}%
	\makeatletter
	\providecommand \@ifxundefined [1]{%
		\@ifx{#1\undefined}
	}%
	\providecommand \@ifnum [1]{%
		\ifnum #1\expandafter \@firstoftwo
		\else \expandafter \@secondoftwo
		\fi
	}%
	\providecommand \@ifx [1]{%
		\ifx #1\expandafter \@firstoftwo
		\else \expandafter \@secondoftwo
		\fi
	}%
	\providecommand \natexlab [1]{#1}%
	\providecommand \enquote  [1]{``#1''}%
	\providecommand \bibnamefont  [1]{#1}%
	\providecommand \bibfnamefont [1]{#1}%
	\providecommand \citenamefont [1]{#1}%
	\providecommand \href@noop [0]{\@secondoftwo}%
	\providecommand \href [0]{\begingroup \@sanitize@url \@href}%
	\providecommand \@href[1]{\@@startlink{#1}\@@href}%
	\providecommand \@@href[1]{\endgroup#1\@@endlink}%
	\providecommand \@sanitize@url [0]{\catcode `\\12\catcode `\$12\catcode
		`\&12\catcode `\#12\catcode `\^12\catcode `\_12\catcode `\%12\relax}%
	\providecommand \@@startlink[1]{}%
	\providecommand \@@endlink[0]{}%
	\providecommand \url  [0]{\begingroup\@sanitize@url \@url }%
	\providecommand \@url [1]{\endgroup\@href {#1}{\urlprefix }}%
	\providecommand \urlprefix  [0]{URL }%
	\providecommand \Eprint [0]{\href }%
	\providecommand \doibase [0]{http://dx.doi.org/}%
	\providecommand \selectlanguage [0]{\@gobble}%
	\providecommand \bibinfo  [0]{\@secondoftwo}%
	\providecommand \bibfield  [0]{\@secondoftwo}%
	\providecommand \translation [1]{[#1]}%
	\providecommand \BibitemOpen [0]{}%
	\providecommand \bibitemStop [0]{}%
	\providecommand \bibitemNoStop [0]{.\EOS\space}%
	\providecommand \EOS [0]{\spacefactor3000\relax}%
	\providecommand \BibitemShut  [1]{\csname bibitem#1\endcsname}%
	\let\auto@bib@innerbib\@empty
	\bibitem [{\citenamefont {Wang}\ \emph {et~al.}(2016)\citenamefont {Wang},
		\citenamefont {Ma},\ and\ \citenamefont {Xue}}]{wang_interface_2016}%
	\BibitemOpen
	\bibfield  {author} {\bibinfo {author} {\bibfnamefont {L.}~\bibnamefont
			{Wang}}, \bibinfo {author} {\bibfnamefont {X.}~\bibnamefont {Ma}}, \ and\
		\bibinfo {author} {\bibfnamefont {Q.-K.}\ \bibnamefont {Xue}},\ }\bibfield
	{title} {\enquote {\bibinfo {title} {Interface high-temperature
				superconductivity},}\ }\href {\doibase 10.1088/0953-2048/29/12/123001}
	{\bibfield  {journal} {\bibinfo  {journal} {Supercond. Sci. Technol.}\
		}\textbf {\bibinfo {volume} {29}},\ \bibinfo {pages} {123001} (\bibinfo
		{year} {2016})}\BibitemShut {NoStop}%
	\bibitem [{\citenamefont {Uchihashi}(2016)}]{uchihashi_two-dimensional_2017}%
	\BibitemOpen
	\bibfield  {author} {\bibinfo {author} {\bibfnamefont {T.}~\bibnamefont
			{Uchihashi}},\ }\bibfield  {title} {\enquote {\bibinfo {title}
			{Two-dimensional superconductors with atomic-scale thickness},}\ }\href
	{\doibase 10.1088/0953-2048/30/1/013002} {\bibfield  {journal} {\bibinfo
			{journal} {Supercond. Sci. Technol.}\ }\textbf {\bibinfo {volume} {30}},\
		\bibinfo {pages} {013002} (\bibinfo {year} {2016})}\BibitemShut {NoStop}%
	\bibitem [{\citenamefont {Xu}\ \emph {et~al.}(2015)\citenamefont {Xu},
		\citenamefont {Wang}, \citenamefont {Liu}, \citenamefont {Chen},
		\citenamefont {Guo}, \citenamefont {Kang}, \citenamefont {Ma}, \citenamefont
		{Cheng},\ and\ \citenamefont {Ren}}]{xu_large-area_2015}%
	\BibitemOpen
	\bibfield  {author} {\bibinfo {author} {\bibfnamefont {C.}~\bibnamefont
			{Xu}}, \bibinfo {author} {\bibfnamefont {L.}~\bibnamefont {Wang}}, \bibinfo
		{author} {\bibfnamefont {Z.}~\bibnamefont {Liu}}, \bibinfo {author}
		{\bibfnamefont {L.}~\bibnamefont {Chen}}, \bibinfo {author} {\bibfnamefont
			{J.}~\bibnamefont {Guo}}, \bibinfo {author} {\bibfnamefont {N.}~\bibnamefont
			{Kang}}, \bibinfo {author} {\bibfnamefont {X.-L.}\ \bibnamefont {Ma}},
		\bibinfo {author} {\bibfnamefont {H.-M.}\ \bibnamefont {Cheng}}, \ and\
		\bibinfo {author} {\bibfnamefont {W.}~\bibnamefont {Ren}},\ }\bibfield
	{title} {\enquote {\bibinfo {title} {Large-area high-quality {{2D}} ultrathin
				{{Mo$_2$C}} superconducting crystals},}\ }\href
	{http://www.nature.com/nmat/journal/vaop/ncurrent/full/nmat4374.html}
	{\bibfield  {journal} {\bibinfo  {journal} {Nat. Mater.}\ }\textbf {\bibinfo
			{volume} {14}},\ \bibinfo {pages} {1135--1141} (\bibinfo {year}
		{2015})}\BibitemShut {NoStop}%
	\bibitem [{\citenamefont {Saito}\ \emph {et~al.}(2016)\citenamefont {Saito},
		\citenamefont {Nojima},\ and\ \citenamefont {Iwasa}}]{saito_highly_2016}%
	\BibitemOpen
	\bibfield  {author} {\bibinfo {author} {\bibfnamefont {Y.}~\bibnamefont
			{Saito}}, \bibinfo {author} {\bibfnamefont {T.}~\bibnamefont {Nojima}}, \
		and\ \bibinfo {author} {\bibfnamefont {Y.}~\bibnamefont {Iwasa}},\ }\bibfield
	{title} {\enquote {\bibinfo {title} {Highly crystalline {{2D}}
				superconductors},}\ }\href {\doibase 10.1038/natrevmats.2016.94} {\bibfield
		{journal} {\bibinfo  {journal} {Nature Reviews Materials}\ }\textbf {\bibinfo
			{volume} {2}},\ \bibinfo {pages} {16094} (\bibinfo {year}
		{2016})}\BibitemShut {NoStop}%
	\bibitem [{\citenamefont {Brun}\ \emph {et~al.}(2017)\citenamefont {Brun},
		\citenamefont {Cren},\ and\ \citenamefont {Roditchev}}]{brun_review_2017}%
	\BibitemOpen
	\bibfield  {author} {\bibinfo {author} {\bibfnamefont {C.}~\bibnamefont
			{Brun}}, \bibinfo {author} {\bibfnamefont {T.}~\bibnamefont {Cren}}, \ and\
		\bibinfo {author} {\bibfnamefont {D.}~\bibnamefont {Roditchev}},\ }\bibfield
	{title} {\enquote {\bibinfo {title} {Review of {{2D}} superconductivity: The
				ultimate case of epitaxial monolayers},}\ }\href {\doibase
		10.1088/0953-2048/30/1/013003} {\bibfield  {journal} {\bibinfo  {journal}
			{Supercond. Sci. Technol.}\ }\textbf {\bibinfo {volume} {30}},\ \bibinfo
		{pages} {013003} (\bibinfo {year} {2017})}\BibitemShut {NoStop}%
	\bibitem [{\citenamefont {Guo}\ \emph {et~al.}(2004)\citenamefont {Guo},
		\citenamefont {Zhang}, \citenamefont {Bao}, \citenamefont {Han},
		\citenamefont {Tang}, \citenamefont {Zhang}, \citenamefont {Zhu},
		\citenamefont {Wang}, \citenamefont {Niu}, \citenamefont {Qiu}, \citenamefont
		{Jia}, \citenamefont {Zhao},\ and\ \citenamefont
		{Xue}}]{guo_superconductivity_2004}%
	\BibitemOpen
	\bibfield  {author} {\bibinfo {author} {\bibfnamefont {Y.}~\bibnamefont
			{Guo}}, \bibinfo {author} {\bibfnamefont {Y.-F.}\ \bibnamefont {Zhang}},
		\bibinfo {author} {\bibfnamefont {X.-Y.}\ \bibnamefont {Bao}}, \bibinfo
		{author} {\bibfnamefont {T.-Z.}\ \bibnamefont {Han}}, \bibinfo {author}
		{\bibfnamefont {Z.}~\bibnamefont {Tang}}, \bibinfo {author} {\bibfnamefont
			{L.-X.}\ \bibnamefont {Zhang}}, \bibinfo {author} {\bibfnamefont {W.-G.}\
			\bibnamefont {Zhu}}, \bibinfo {author} {\bibfnamefont {E.~G.}\ \bibnamefont
			{Wang}}, \bibinfo {author} {\bibfnamefont {Q.}~\bibnamefont {Niu}}, \bibinfo
		{author} {\bibfnamefont {Z.~Q.}\ \bibnamefont {Qiu}}, \bibinfo {author}
		{\bibfnamefont {J.-F.}\ \bibnamefont {Jia}}, \bibinfo {author} {\bibfnamefont
			{Z.-X.}\ \bibnamefont {Zhao}}, \ and\ \bibinfo {author} {\bibfnamefont
			{Q.-K.}\ \bibnamefont {Xue}},\ }\bibfield  {title} {\enquote {\bibinfo
			{title} {Superconductivity {Modulated} by {Quantum} {Size} {Effects}},}\
	}\href {\doibase 10.1126/science.1105130} {\bibfield  {journal} {\bibinfo
			{journal} {Science}\ }\textbf {\bibinfo {volume} {306}},\ \bibinfo {pages}
		{1915--1917} (\bibinfo {year} {2004})}\BibitemShut {NoStop}%
	\bibitem [{\citenamefont {\"{O}zer}\ \emph {et~al.}(2006)\citenamefont
		{\"{O}zer}, \citenamefont {Thompson},\ and\ \citenamefont
		{Weitering}}]{ozer_hard_2006}%
	\BibitemOpen
	\bibfield  {author} {\bibinfo {author} {\bibfnamefont {M.~M.}\ \bibnamefont
			{\"{O}zer}}, \bibinfo {author} {\bibfnamefont {J.~R.}\ \bibnamefont
			{Thompson}}, \ and\ \bibinfo {author} {\bibfnamefont {H.~H.}\ \bibnamefont
			{Weitering}},\ }\bibfield  {title} {\enquote {\bibinfo {title} {Hard
				superconductivity of a soft metal in the quantum regime},}\ }\href {\doibase
		10.1038/nphys244} {\bibfield  {journal} {\bibinfo  {journal} {Nat. Phys.}\
		}\textbf {\bibinfo {volume} {2}},\ \bibinfo {pages} {173--176} (\bibinfo
		{year} {2006})}\BibitemShut {NoStop}%
	\bibitem [{\citenamefont {Zhang}\ \emph {et~al.}(2010)\citenamefont {Zhang},
		\citenamefont {Cheng}, \citenamefont {Li}, \citenamefont {Sun}, \citenamefont
		{Wang}, \citenamefont {Zhu}, \citenamefont {He}, \citenamefont {Wang},
		\citenamefont {Ma}, \citenamefont {Chen}, \citenamefont {Wang}, \citenamefont
		{Liu}, \citenamefont {Lin}, \citenamefont {Jia},\ and\ \citenamefont
		{Xue}}]{zhang_superconductivity_2010}%
	\BibitemOpen
	\bibfield  {author} {\bibinfo {author} {\bibfnamefont {T.}~\bibnamefont
			{Zhang}}, \bibinfo {author} {\bibfnamefont {P.}~\bibnamefont {Cheng}},
		\bibinfo {author} {\bibfnamefont {W.-J.}\ \bibnamefont {Li}}, \bibinfo
		{author} {\bibfnamefont {Y.-J.}\ \bibnamefont {Sun}}, \bibinfo {author}
		{\bibfnamefont {G.}~\bibnamefont {Wang}}, \bibinfo {author} {\bibfnamefont
			{X.-G.}\ \bibnamefont {Zhu}}, \bibinfo {author} {\bibfnamefont
			{K.}~\bibnamefont {He}}, \bibinfo {author} {\bibfnamefont {L.}~\bibnamefont
			{Wang}}, \bibinfo {author} {\bibfnamefont {X.}~\bibnamefont {Ma}}, \bibinfo
		{author} {\bibfnamefont {X.}~\bibnamefont {Chen}}, \bibinfo {author}
		{\bibfnamefont {Y.}~\bibnamefont {Wang}}, \bibinfo {author} {\bibfnamefont
			{Y.}~\bibnamefont {Liu}}, \bibinfo {author} {\bibfnamefont {H.-Q.}\
			\bibnamefont {Lin}}, \bibinfo {author} {\bibfnamefont {J.-F.}\ \bibnamefont
			{Jia}}, \ and\ \bibinfo {author} {\bibfnamefont {Q.-K.}\ \bibnamefont
			{Xue}},\ }\bibfield  {title} {\enquote {\bibinfo {title} {Superconductivity
				in one-atomic-layer metal films grown on {Si}(111)},}\ }\href {\doibase
		10.1038/nphys1499} {\bibfield  {journal} {\bibinfo  {journal} {Nat. Phys.}\
		}\textbf {\bibinfo {volume} {6}},\ \bibinfo {pages} {104--108} (\bibinfo
		{year} {2010})}\BibitemShut {NoStop}%
	\bibitem [{\citenamefont {St\c{e}pniak}\ \emph {et~al.}(2014)\citenamefont
		{St\c{e}pniak}, \citenamefont {Leon~Vanegas}, \citenamefont {Caminale},
		\citenamefont {Oka}, \citenamefont {Sander},\ and\ \citenamefont
		{Kirschner}}]{stepniak_atomic_2014}%
	\BibitemOpen
	\bibfield  {author} {\bibinfo {author} {\bibfnamefont {A.}~\bibnamefont
			{St\c{e}pniak}}, \bibinfo {author} {\bibfnamefont {A.}~\bibnamefont
			{Leon~Vanegas}}, \bibinfo {author} {\bibfnamefont {M.}~\bibnamefont
			{Caminale}}, \bibinfo {author} {\bibfnamefont {H.}~\bibnamefont {Oka}},
		\bibinfo {author} {\bibfnamefont {D.}~\bibnamefont {Sander}}, \ and\ \bibinfo
		{author} {\bibfnamefont {J.}~\bibnamefont {Kirschner}},\ }\bibfield  {title}
	{\enquote {\bibinfo {title} {Atomic layer superconductivity},}\ }\href
	{\doibase 10.1002/sia.5516} {\bibfield  {journal} {\bibinfo  {journal} {Surf.
				Interface Anal.}\ }\textbf {\bibinfo {volume} {46}},\ \bibinfo {pages}
		{1262--1267} (\bibinfo {year} {2014})}\BibitemShut {NoStop}%
	\bibitem [{\citenamefont {Zhang}\ \emph
		{et~al.}(2015{\natexlab{a}})\citenamefont {Zhang}, \citenamefont {Sun},
		\citenamefont {Li}, \citenamefont {Peng}, \citenamefont {Song}, \citenamefont
		{Xing}, \citenamefont {Zhang}, \citenamefont {Guan}, \citenamefont {Li},
		\citenamefont {Zhao}, \citenamefont {Ji}, \citenamefont {Wang}, \citenamefont
		{He}, \citenamefont {Chen}, \citenamefont {Gu}, \citenamefont {Ling},
		\citenamefont {Tian}, \citenamefont {Li}, \citenamefont {Xie}, \citenamefont
		{Liu}, \citenamefont {Yang}, \citenamefont {Xue}, \citenamefont {Wang},\ and\
		\citenamefont {Ma}}]{zhang_detection_2015}%
	\BibitemOpen
	\bibfield  {author} {\bibinfo {author} {\bibfnamefont {H.-M.}\ \bibnamefont
			{Zhang}}, \bibinfo {author} {\bibfnamefont {Y.}~\bibnamefont {Sun}}, \bibinfo
		{author} {\bibfnamefont {W.}~\bibnamefont {Li}}, \bibinfo {author}
		{\bibfnamefont {J.-P.}\ \bibnamefont {Peng}}, \bibinfo {author}
		{\bibfnamefont {C.-L.}\ \bibnamefont {Song}}, \bibinfo {author}
		{\bibfnamefont {Y.}~\bibnamefont {Xing}}, \bibinfo {author} {\bibfnamefont
			{Q.}~\bibnamefont {Zhang}}, \bibinfo {author} {\bibfnamefont
			{J.}~\bibnamefont {Guan}}, \bibinfo {author} {\bibfnamefont {Z.}~\bibnamefont
			{Li}}, \bibinfo {author} {\bibfnamefont {Y.}~\bibnamefont {Zhao}}, \bibinfo
		{author} {\bibfnamefont {S.}~\bibnamefont {Ji}}, \bibinfo {author}
		{\bibfnamefont {L.}~\bibnamefont {Wang}}, \bibinfo {author} {\bibfnamefont
			{K.}~\bibnamefont {He}}, \bibinfo {author} {\bibfnamefont {X.}~\bibnamefont
			{Chen}}, \bibinfo {author} {\bibfnamefont {L.}~\bibnamefont {Gu}}, \bibinfo
		{author} {\bibfnamefont {L.}~\bibnamefont {Ling}}, \bibinfo {author}
		{\bibfnamefont {M.}~\bibnamefont {Tian}}, \bibinfo {author} {\bibfnamefont
			{L.}~\bibnamefont {Li}}, \bibinfo {author} {\bibfnamefont {X.~C.}\
			\bibnamefont {Xie}}, \bibinfo {author} {\bibfnamefont {J.}~\bibnamefont
			{Liu}}, \bibinfo {author} {\bibfnamefont {H.}~\bibnamefont {Yang}}, \bibinfo
		{author} {\bibfnamefont {Q.-K.}\ \bibnamefont {Xue}}, \bibinfo {author}
		{\bibfnamefont {J.}~\bibnamefont {Wang}}, \ and\ \bibinfo {author}
		{\bibfnamefont {X.}~\bibnamefont {Ma}},\ }\bibfield  {title} {\enquote
		{\bibinfo {title} {Detection of a {{Superconducting Phase}} in a
				{{Two}}-{{Atom Layer}} of {{Hexagonal Ga Film Grown}} on {{Semiconducting
						GaN}}(0001)},}\ }\href {\doibase 10.1103/PhysRevLett.114.107003} {\bibfield
		{journal} {\bibinfo  {journal} {Phys. Rev. Lett.}\ }\textbf {\bibinfo
			{volume} {114}},\ \bibinfo {pages} {107003} (\bibinfo {year}
		{2015}{\natexlab{a}})}\BibitemShut {NoStop}%
	\bibitem [{\citenamefont {Goldman}\ and\ \citenamefont
		{Markovi\'{c}}(2008)}]{goldman_superconductorinsulator_2008}%
	\BibitemOpen
	\bibfield  {author} {\bibinfo {author} {\bibfnamefont {A.~M.}\ \bibnamefont
			{Goldman}}\ and\ \bibinfo {author} {\bibfnamefont {N.}~\bibnamefont
			{Markovi\'{c}}},\ }\bibfield  {title} {\enquote {\bibinfo {title}
			{Superconductor\sloppy --{Insulator} {Transitions} in the
				{Two}\sloppy-‐{Dimensional} {Limit}},}\ }\href {\doibase 10.1063/1.882069}
	{\bibfield  {journal} {\bibinfo  {journal} {Phys. Today}\ }\textbf {\bibinfo
			{volume} {51}},\ \bibinfo {pages} {39--44} (\bibinfo {year}
		{2008})}\BibitemShut {NoStop}%
	\bibitem [{\citenamefont {Dubi}\ \emph {et~al.}(2007)\citenamefont {Dubi},
		\citenamefont {Meir},\ and\ \citenamefont {Avishai}}]{dubi_nature_2007}%
	\BibitemOpen
	\bibfield  {author} {\bibinfo {author} {\bibfnamefont {Y.}~\bibnamefont
			{Dubi}}, \bibinfo {author} {\bibfnamefont {Y.}~\bibnamefont {Meir}}, \ and\
		\bibinfo {author} {\bibfnamefont {Y.}~\bibnamefont {Avishai}},\ }\bibfield
	{title} {\enquote {\bibinfo {title} {Nature of the
				superconductor\sloppy--insulator transition in disordered superconductors},}\
	}\href {\doibase 10.1038/nature06180} {\bibfield  {journal} {\bibinfo
			{journal} {Nature (London)}\ }\textbf {\bibinfo {volume} {449}},\ \bibinfo
		{pages} {876--880} (\bibinfo {year} {2007})}\BibitemShut {NoStop}%
	\bibitem [{\citenamefont {Ge}\ \emph {et~al.}(2015)\citenamefont {Ge},
		\citenamefont {Liu}, \citenamefont {Liu}, \citenamefont {Gao}, \citenamefont
		{Qian}, \citenamefont {Xue}, \citenamefont {Liu},\ and\ \citenamefont
		{Jia}}]{ge_superconductivity_2015}%
	\BibitemOpen
	\bibfield  {author} {\bibinfo {author} {\bibfnamefont {J.-F.}\ \bibnamefont
			{Ge}}, \bibinfo {author} {\bibfnamefont {Z.-L.}\ \bibnamefont {Liu}},
		\bibinfo {author} {\bibfnamefont {C.}~\bibnamefont {Liu}}, \bibinfo {author}
		{\bibfnamefont {C.-L.}\ \bibnamefont {Gao}}, \bibinfo {author} {\bibfnamefont
			{D.}~\bibnamefont {Qian}}, \bibinfo {author} {\bibfnamefont {Q.-K.}\
			\bibnamefont {Xue}}, \bibinfo {author} {\bibfnamefont {Y.}~\bibnamefont
			{Liu}}, \ and\ \bibinfo {author} {\bibfnamefont {J.-F.}\ \bibnamefont
			{Jia}},\ }\bibfield  {title} {\enquote {\bibinfo {title} {Superconductivity
				above 100 {K} in single-layer {FeSe} films on doped {SrTiO}$_3$},}\ }\href
	{\doibase 10.1038/nmat4153} {\bibfield  {journal} {\bibinfo  {journal} {Nat.
				Mater.}\ }\textbf {\bibinfo {volume} {14}},\ \bibinfo {pages} {285--289}
		(\bibinfo {year} {2015})}\BibitemShut {NoStop}%
	\bibitem [{\citenamefont {Xi}\ \emph {et~al.}(2016)\citenamefont {Xi},
		\citenamefont {Wang}, \citenamefont {Zhao}, \citenamefont {Park},
		\citenamefont {Law}, \citenamefont {Berger}, \citenamefont {Forr\'{o}},
		\citenamefont {Shan},\ and\ \citenamefont {Mak}}]{xi_ising_2016}%
	\BibitemOpen
	\bibfield  {author} {\bibinfo {author} {\bibfnamefont {X.}~\bibnamefont
			{Xi}}, \bibinfo {author} {\bibfnamefont {Z.}~\bibnamefont {Wang}}, \bibinfo
		{author} {\bibfnamefont {W.}~\bibnamefont {Zhao}}, \bibinfo {author}
		{\bibfnamefont {J.-H.}\ \bibnamefont {Park}}, \bibinfo {author}
		{\bibfnamefont {K.~T.}\ \bibnamefont {Law}}, \bibinfo {author} {\bibfnamefont
			{H.}~\bibnamefont {Berger}}, \bibinfo {author} {\bibfnamefont
			{L.}~\bibnamefont {Forr\'{o}}}, \bibinfo {author} {\bibfnamefont
			{J.}~\bibnamefont {Shan}}, \ and\ \bibinfo {author} {\bibfnamefont {K.~F.}\
			\bibnamefont {Mak}},\ }\bibfield  {title} {\enquote {\bibinfo {title} {Ising
				pairing in superconducting {NbSe}{$_2$} atomic layers},}\ }\href {\doibase
		10.1038/nphys3538} {\bibfield  {journal} {\bibinfo  {journal} {Nat. Phys.}\
		}\textbf {\bibinfo {volume} {12}},\ \bibinfo {pages} {139--143} (\bibinfo
		{year} {2016})}\BibitemShut {NoStop}%
	\bibitem [{\citenamefont {Shanenko}\ \emph {et~al.}(2006)\citenamefont
		{Shanenko}, \citenamefont {Croitoru}, \citenamefont {Zgirski}, \citenamefont
		{Peeters},\ and\ \citenamefont {Arutyunov}}]{shanenko_size-dependent_2006}%
	\BibitemOpen
	\bibfield  {author} {\bibinfo {author} {\bibfnamefont {A.~A.}\ \bibnamefont
			{Shanenko}}, \bibinfo {author} {\bibfnamefont {M.~D.}\ \bibnamefont
			{Croitoru}}, \bibinfo {author} {\bibfnamefont {M.}~\bibnamefont {Zgirski}},
		\bibinfo {author} {\bibfnamefont {F.~M.}\ \bibnamefont {Peeters}}, \ and\
		\bibinfo {author} {\bibfnamefont {K.}~\bibnamefont {Arutyunov}},\ }\bibfield
	{title} {\enquote {\bibinfo {title} {Size-dependent enhancement of
				superconductivity in {{Al}} and {{Sn}} nanowires: {{Shape}}-resonance
				effect},}\ }\href {\doibase 10.1103/PhysRevB.74.052502} {\bibfield  {journal}
		{\bibinfo  {journal} {Phys. Rev. B}\ }\textbf {\bibinfo {volume} {74}},\
		\bibinfo {pages} {052502} (\bibinfo {year} {2006})}\BibitemShut {NoStop}%
	\bibitem [{\citenamefont {Croitoru}\ \emph {et~al.}(2007)\citenamefont
		{Croitoru}, \citenamefont {Shanenko},\ and\ \citenamefont
		{Peeters}}]{croitoru_dependence_2007}%
	\BibitemOpen
	\bibfield  {author} {\bibinfo {author} {\bibfnamefont {M.~D.}\ \bibnamefont
			{Croitoru}}, \bibinfo {author} {\bibfnamefont {A.~A.}\ \bibnamefont
			{Shanenko}}, \ and\ \bibinfo {author} {\bibfnamefont {F.~M.}\ \bibnamefont
			{Peeters}},\ }\bibfield  {title} {\enquote {\bibinfo {title} {Dependence of
				superconducting properties on the size and shape of a nanoscale
				superconductor: {{From}} nanowire to film},}\ }\href {\doibase
		10.1103/PhysRevB.76.024511} {\bibfield  {journal} {\bibinfo  {journal} {Phys.
				Rev. B}\ }\textbf {\bibinfo {volume} {76}},\ \bibinfo {pages} {024511}
		(\bibinfo {year} {2007})}\BibitemShut {NoStop}%
	\bibitem [{\citenamefont {Romero-Berm\'{u}dez}\ and\ \citenamefont
		{Garc\'{i}a-Garc\'{i}a}(2014{\natexlab{a}})}]{romero-bermudez_shape_2014}%
	\BibitemOpen
	\bibfield  {author} {\bibinfo {author} {\bibfnamefont {A.}~\bibnamefont
			{Romero-Berm\'{u}dez}}\ and\ \bibinfo {author} {\bibfnamefont {A.~M.}\
			\bibnamefont {Garc\'{i}a-Garc\'{i}a}},\ }\bibfield  {title} {\enquote
		{\bibinfo {title} {Shape resonances and shell effects in thin-film multiband
				superconductors},}\ }\href {\doibase 10.1103/PhysRevB.89.024510} {\bibfield
		{journal} {\bibinfo  {journal} {Phys. Rev. B}\ }\textbf {\bibinfo {volume}
			{89}},\ \bibinfo {pages} {024510} (\bibinfo {year}
		{2014}{\natexlab{a}})}\BibitemShut {NoStop}%
	\bibitem [{\citenamefont {Romero-Berm\'{u}dez}\ and\ \citenamefont
		{Garc\'{i}a-Garc\'{i}a}(2014{\natexlab{b}})}]{romero-bermudez_size_2014}%
	\BibitemOpen
	\bibfield  {author} {\bibinfo {author} {\bibfnamefont {A.}~\bibnamefont
			{Romero-Berm\'{u}dez}}\ and\ \bibinfo {author} {\bibfnamefont {A.~M.}\
			\bibnamefont {Garc\'{i}a-Garc\'{i}a}},\ }\bibfield  {title} {\enquote
		{\bibinfo {title} {Size effects in superconducting thin films coupled to a
				substrate},}\ }\href {\doibase 10.1103/PhysRevB.89.064508} {\bibfield
		{journal} {\bibinfo  {journal} {Phys. Rev. B}\ }\textbf {\bibinfo {volume}
			{89}},\ \bibinfo {pages} {064508} (\bibinfo {year}
		{2014}{\natexlab{b}})}\BibitemShut {NoStop}%
	\bibitem [{\citenamefont {Chen}\ \emph
		{et~al.}(2012{\natexlab{a}})\citenamefont {Chen}, \citenamefont {Shanenko},
		\citenamefont {Perali},\ and\ \citenamefont
		{Peeters}}]{perali_molecule-like_2012}%
	\BibitemOpen
	\bibfield  {author} {\bibinfo {author} {\bibfnamefont {Y.}~\bibnamefont
			{Chen}}, \bibinfo {author} {\bibfnamefont {A.~A.}\ \bibnamefont {Shanenko}},
		\bibinfo {author} {\bibfnamefont {A.}~\bibnamefont {Perali}}, \ and\ \bibinfo
		{author} {\bibfnamefont {F.~M.}\ \bibnamefont {Peeters}},\ }\bibfield
	{title} {\enquote {\bibinfo {title} {Superconducting nanofilms: molecule-like
				pairing induced by quantum confinement},}\ }\href {\doibase
		10.1088/0953-8984/24/18/185701} {\bibfield  {journal} {\bibinfo  {journal}
			{J. Phys.: Condens. Matter}\ }\textbf {\bibinfo {volume} {24}},\ \bibinfo
		{pages} {185701} (\bibinfo {year} {2012}{\natexlab{a}})}\BibitemShut
	{NoStop}%
	\bibitem [{\citenamefont {Bianconi}\ \emph {et~al.}(2014)\citenamefont
		{Bianconi}, \citenamefont {Innocenti}, \citenamefont {Valletta},\ and\
		\citenamefont {Perali}}]{bianconi_shape_resonance_2014}%
	\BibitemOpen
	\bibfield  {author} {\bibinfo {author} {\bibfnamefont {A.}~\bibnamefont
			{Bianconi}}, \bibinfo {author} {\bibfnamefont {D.}~\bibnamefont {Innocenti}},
		\bibinfo {author} {\bibfnamefont {A.}~\bibnamefont {Valletta}}, \ and\
		\bibinfo {author} {\bibfnamefont {A.}~\bibnamefont {Perali}},\ }\bibfield
	{title} {\enquote {\bibinfo {title} {Shape {Resonances} in superconducting
				gaps in a {2DEG} at oxide-oxide interface},}\ }\href {\doibase
		10.1088/1742-6596/529/1/012007} {\bibfield  {journal} {\bibinfo  {journal}
			{J. Phys.: Conf. Ser.}\ }\textbf {\bibinfo {volume} {529}},\ \bibinfo {pages}
		{012007} (\bibinfo {year} {2014})}\BibitemShut {NoStop}%
	\bibitem [{\citenamefont {Innocenti}\ \emph {et~al.}(2010)\citenamefont
		{Innocenti}, \citenamefont {Poccia}, \citenamefont {Ricci}, \citenamefont
		{Valletta}, \citenamefont {Caprara}, \citenamefont {Perali},\ and\
		\citenamefont {Bianconi}}]{innocenti_resonant_2010}%
	\BibitemOpen
	\bibfield  {author} {\bibinfo {author} {\bibfnamefont {D.}~\bibnamefont
			{Innocenti}}, \bibinfo {author} {\bibfnamefont {N.}~\bibnamefont {Poccia}},
		\bibinfo {author} {\bibfnamefont {A.}~\bibnamefont {Ricci}}, \bibinfo
		{author} {\bibfnamefont {A.}~\bibnamefont {Valletta}}, \bibinfo {author}
		{\bibfnamefont {S.}~\bibnamefont {Caprara}}, \bibinfo {author} {\bibfnamefont
			{A.}~\bibnamefont {Perali}}, \ and\ \bibinfo {author} {\bibfnamefont
			{A.}~\bibnamefont {Bianconi}},\ }\bibfield  {title} {\enquote {\bibinfo
			{title} {Resonant and crossover phenomena in a multiband superconductor:
				{{Tuning}} the chemical potential near a band edge},}\ }\href {\doibase
		10.1103/PhysRevB.82.184528} {\bibfield  {journal} {\bibinfo  {journal} {Phys.
				Rev. B}\ }\textbf {\bibinfo {volume} {82}},\ \bibinfo {pages} {184528}
		(\bibinfo {year} {2010})}\BibitemShut {NoStop}%
	\bibitem [{\citenamefont {Doria}\ \emph {et~al.}(2016)\citenamefont {Doria},
		\citenamefont {Cariglia},\ and\ \citenamefont
		{Perali}}]{doria_multigap_2016}%
	\BibitemOpen
	\bibfield  {author} {\bibinfo {author} {\bibfnamefont {M.~M.}\ \bibnamefont
			{Doria}}, \bibinfo {author} {\bibfnamefont {M.}~\bibnamefont {Cariglia}}, \
		and\ \bibinfo {author} {\bibfnamefont {A.}~\bibnamefont {Perali}},\
	}\bibfield  {title} {\enquote {\bibinfo {title} {Multigap superconductivity
				and barrier-driven resonances in superconducting nanofilms with an inner
				potential barrier},}\ }\href {\doibase 10.1103/PhysRevB.94.224513} {\bibfield
		{journal} {\bibinfo  {journal} {Phys. Rev. B}\ }\textbf {\bibinfo {volume}
			{94}},\ \bibinfo {pages} {224513} (\bibinfo {year} {2016})}\BibitemShut
	{NoStop}%
	\bibitem [{\citenamefont {Brun}\ \emph {et~al.}(2014)\citenamefont {Brun},
		\citenamefont {Cren}, \citenamefont {Cherkez}, \citenamefont {Debontridder},
		\citenamefont {Pons}, \citenamefont {Fokin}, \citenamefont {Tringides},
		\citenamefont {Bozhko}, \citenamefont {Ioffe}, \citenamefont {Altshuler},\
		and\ \citenamefont {Roditchev}}]{brun_remarkable_2014}%
	\BibitemOpen
	\bibfield  {author} {\bibinfo {author} {\bibfnamefont {C.}~\bibnamefont
			{Brun}}, \bibinfo {author} {\bibfnamefont {T.}~\bibnamefont {Cren}}, \bibinfo
		{author} {\bibfnamefont {V.}~\bibnamefont {Cherkez}}, \bibinfo {author}
		{\bibfnamefont {F.}~\bibnamefont {Debontridder}}, \bibinfo {author}
		{\bibfnamefont {S.}~\bibnamefont {Pons}}, \bibinfo {author} {\bibfnamefont
			{D.}~\bibnamefont {Fokin}}, \bibinfo {author} {\bibfnamefont {M.~C.}\
			\bibnamefont {Tringides}}, \bibinfo {author} {\bibfnamefont {S.}~\bibnamefont
			{Bozhko}}, \bibinfo {author} {\bibfnamefont {L.~B.}\ \bibnamefont {Ioffe}},
		\bibinfo {author} {\bibfnamefont {B.~L.}\ \bibnamefont {Altshuler}}, \ and\
		\bibinfo {author} {\bibfnamefont {D.}~\bibnamefont {Roditchev}},\ }\bibfield
	{title} {\enquote {\bibinfo {title} {Remarkable {Effects} of {Disorder} on
				{Superconductivity} of {Single} {Atomic} {Layers} of {Lead} on {Silicon}},}\
	}\href {\doibase 10.1038/nphys2937} {\bibfield  {journal} {\bibinfo
			{journal} {Nat. Phys.}\ }\textbf {\bibinfo {volume} {10}},\ \bibinfo {pages}
		{444--450} (\bibinfo {year} {2014})}\BibitemShut {NoStop}%
	\bibitem [{\citenamefont {Yoshizawa}\ \emph {et~al.}(2014)\citenamefont
		{Yoshizawa}, \citenamefont {Kim}, \citenamefont {Kawakami}, \citenamefont
		{Nagai}, \citenamefont {Nakayama}, \citenamefont {Hu}, \citenamefont
		{Hasegawa},\ and\ \citenamefont {Uchihashi}}]{yoshizawa_imaging_2014}%
	\BibitemOpen
	\bibfield  {author} {\bibinfo {author} {\bibfnamefont {S.}~\bibnamefont
			{Yoshizawa}}, \bibinfo {author} {\bibfnamefont {H.}~\bibnamefont {Kim}},
		\bibinfo {author} {\bibfnamefont {T.}~\bibnamefont {Kawakami}}, \bibinfo
		{author} {\bibfnamefont {Y.}~\bibnamefont {Nagai}}, \bibinfo {author}
		{\bibfnamefont {T.}~\bibnamefont {Nakayama}}, \bibinfo {author}
		{\bibfnamefont {X.}~\bibnamefont {Hu}}, \bibinfo {author} {\bibfnamefont
			{Y.}~\bibnamefont {Hasegawa}}, \ and\ \bibinfo {author} {\bibfnamefont
			{T.}~\bibnamefont {Uchihashi}},\ }\bibfield  {title} {\enquote {\bibinfo
			{title} {Imaging {Josephson} {Vortices} on the {Surface} {Superconductor}
				$\mathrm{Si}(111)\text{-}(\sqrt{7}\ifmmode\times\else\texttimes\fi{}\sqrt{3})\text{-}\mathrm{In}$
				using a {Scanning} {Tunneling} {Microscope}},}\ }\href {\doibase
		10.1103/PhysRevLett.113.247004} {\bibfield  {journal} {\bibinfo  {journal}
			{Phys. Rev. Lett.}\ }\textbf {\bibinfo {volume} {113}},\ \bibinfo {pages}
		{247004} (\bibinfo {year} {2014})}\BibitemShut {NoStop}%
	\bibitem [{\citenamefont {Kim}\ \emph {et~al.}(2016)\citenamefont {Kim},
		\citenamefont {Lin}, \citenamefont {Graf}, \citenamefont {Miyata},
		\citenamefont {Nagai}, \citenamefont {Kato},\ and\ \citenamefont
		{Hasegawa}}]{kim_electrical_2016}%
	\BibitemOpen
	\bibfield  {author} {\bibinfo {author} {\bibfnamefont {H.}~\bibnamefont
			{Kim}}, \bibinfo {author} {\bibfnamefont {S.-Z.}\ \bibnamefont {Lin}},
		\bibinfo {author} {\bibfnamefont {M.~J.}\ \bibnamefont {Graf}}, \bibinfo
		{author} {\bibfnamefont {Y.}~\bibnamefont {Miyata}}, \bibinfo {author}
		{\bibfnamefont {Y.}~\bibnamefont {Nagai}}, \bibinfo {author} {\bibfnamefont
			{T.}~\bibnamefont {Kato}}, \ and\ \bibinfo {author} {\bibfnamefont
			{Y.}~\bibnamefont {Hasegawa}},\ }\bibfield  {title} {\enquote {\bibinfo
			{title} {Electrical {Conductivity} through a {Single} {Atomic} {Step}
				{Measured} with the {Proximity}-{Induced} {Superconducting} {Pair}
				{Correlation}},}\ }\href {\doibase 10.1103/PhysRevLett.117.116802} {\bibfield
		{journal} {\bibinfo  {journal} {Phys. Rev. Lett.}\ }\textbf {\bibinfo
			{volume} {117}},\ \bibinfo {pages} {116802} (\bibinfo {year}
		{2016})}\BibitemShut {NoStop}%
	\bibitem [{\citenamefont {Zhang}\ \emph {et~al.}(2012)\citenamefont {Zhang},
		\citenamefont {Covaci}, \citenamefont {Milo\ifmmode \check{s}\else
			\v{s}\fi{}evi\ifmmode~\acute{c}\else \'{c}\fi{}}, \citenamefont
		{Berdiyorov},\ and\ \citenamefont {Peeters}}]{zhang_unconventional_2012}%
	\BibitemOpen
	\bibfield  {author} {\bibinfo {author} {\bibfnamefont {L.-F.}\ \bibnamefont
			{Zhang}}, \bibinfo {author} {\bibfnamefont {L.}~\bibnamefont {Covaci}},
		\bibinfo {author} {\bibfnamefont {M.~V.}\ \bibnamefont {Milo\ifmmode
				\check{s}\else \v{s}\fi{}evi\ifmmode~\acute{c}\else \'{c}\fi{}}}, \bibinfo
		{author} {\bibfnamefont {G.~R.}\ \bibnamefont {Berdiyorov}}, \ and\ \bibinfo
		{author} {\bibfnamefont {F.~M.}\ \bibnamefont {Peeters}},\ }\bibfield
	{title} {\enquote {\bibinfo {title} {Unconventional {Vortex} {States} in
				{Nanoscale} {Superconductors} {Due} to {Shape}-{Induced} {Resonances} in the
				{Inhomogeneous} {Cooper}-{pair} {Condensate}},}\ }\href {\doibase
		10.1103/PhysRevLett.109.107001} {\bibfield  {journal} {\bibinfo  {journal}
			{Phys. Rev. Lett.}\ }\textbf {\bibinfo {volume} {109}},\ \bibinfo {pages}
		{107001} (\bibinfo {year} {2012})}\BibitemShut {NoStop}%
	\bibitem [{\citenamefont {Shanenko}\ \emph {et~al.}(2007)\citenamefont
		{Shanenko}, \citenamefont {Croitoru}, \citenamefont {Mints},\ and\
		\citenamefont {Peeters}}]{shanenko_new_2007}%
	\BibitemOpen
	\bibfield  {author} {\bibinfo {author} {\bibfnamefont {A.~A.}\ \bibnamefont
			{Shanenko}}, \bibinfo {author} {\bibfnamefont {M.~D.}\ \bibnamefont
			{Croitoru}}, \bibinfo {author} {\bibfnamefont {R.~G.}\ \bibnamefont {Mints}},
		\ and\ \bibinfo {author} {\bibfnamefont {F.~M.}\ \bibnamefont {Peeters}},\
	}\bibfield  {title} {\enquote {\bibinfo {title} {New {Andreev}-{Type}
				{States} in {Superconducting} {Nanowires}},}\ }\href {\doibase
		10.1103/PhysRevLett.99.067007} {\bibfield  {journal} {\bibinfo  {journal}
			{Phys. Rev. Lett.}\ }\textbf {\bibinfo {volume} {99}},\ \bibinfo {pages}
		{067007} (\bibinfo {year} {2007})}\BibitemShut {NoStop}%
	\bibitem [{\citenamefont {Zhang}\ \emph
		{et~al.}(2015{\natexlab{b}})\citenamefont {Zhang}, \citenamefont {Covaci},\
		and\ \citenamefont {Peeters}}]{zhang_position-dependent_2015}%
	\BibitemOpen
	\bibfield  {author} {\bibinfo {author} {\bibfnamefont {L.-F.}\ \bibnamefont
			{Zhang}}, \bibinfo {author} {\bibfnamefont {L.}~\bibnamefont {Covaci}}, \
		and\ \bibinfo {author} {\bibfnamefont {F.~M.}\ \bibnamefont {Peeters}},\
	}\bibfield  {title} {\enquote {\bibinfo {title} {Position-dependent effect of
				non-magnetic impurities on superconducting properties of nanowires},}\ }\href
	{\doibase 10.1209/0295-5075/109/17010} {\bibfield  {journal} {\bibinfo
			{journal} {Europhys. Lett.}\ }\textbf {\bibinfo {volume} {109}},\ \bibinfo
		{pages} {17010} (\bibinfo {year} {2015}{\natexlab{b}})}\BibitemShut {NoStop}%
	\bibitem [{\citenamefont {Zhang}\ \emph
		{et~al.}(2015{\natexlab{c}})\citenamefont {Zhang}, \citenamefont {Covaci},\
		and\ \citenamefont {Peeters}}]{zhang_tomasch_2015}%
	\BibitemOpen
	\bibfield  {author} {\bibinfo {author} {\bibfnamefont {L.-F.}\ \bibnamefont
			{Zhang}}, \bibinfo {author} {\bibfnamefont {L.}~\bibnamefont {Covaci}}, \
		and\ \bibinfo {author} {\bibfnamefont {F.~M.}\ \bibnamefont {Peeters}},\
	}\bibfield  {title} {\enquote {\bibinfo {title} {Tomasch effect in nanoscale
				superconductors},}\ }\href {\doibase 10.1103/PhysRevB.91.024508} {\bibfield
		{journal} {\bibinfo  {journal} {Phys. Rev. B}\ }\textbf {\bibinfo {volume}
			{91}},\ \bibinfo {pages} {024508} (\bibinfo {year}
		{2015}{\natexlab{c}})}\BibitemShut {NoStop}%
	\bibitem [{\citenamefont {Bao}\ \emph {et~al.}(2005)\citenamefont {Bao},
		\citenamefont {Zhang}, \citenamefont {Wang}, \citenamefont {Jia},
		\citenamefont {Xue}, \citenamefont {Xie},\ and\ \citenamefont
		{Zhao}}]{bao_quantum_2005}%
	\BibitemOpen
	\bibfield  {author} {\bibinfo {author} {\bibfnamefont {X.-Y.}\ \bibnamefont
			{Bao}}, \bibinfo {author} {\bibfnamefont {Y.-F.}\ \bibnamefont {Zhang}},
		\bibinfo {author} {\bibfnamefont {Y.}~\bibnamefont {Wang}}, \bibinfo {author}
		{\bibfnamefont {J.-F.}\ \bibnamefont {Jia}}, \bibinfo {author} {\bibfnamefont
			{Q.-K.}\ \bibnamefont {Xue}}, \bibinfo {author} {\bibfnamefont {X.~C.}\
			\bibnamefont {Xie}}, \ and\ \bibinfo {author} {\bibfnamefont {Z.-X.}\
			\bibnamefont {Zhao}},\ }\bibfield  {title} {\enquote {\bibinfo {title}
			{Quantum size effects on the perpendicular upper critical field in ultrathin
				lead films},}\ }\href {\doibase 10.1103/PhysRevLett.95.247005} {\bibfield
		{journal} {\bibinfo  {journal} {Phys. Rev. Lett.}\ }\textbf {\bibinfo
			{volume} {95}},\ \bibinfo {pages} {247005} (\bibinfo {year}
		{2005})}\BibitemShut {NoStop}%
	\bibitem [{\citenamefont {Eom}\ \emph {et~al.}(2006)\citenamefont {Eom},
		\citenamefont {Qin}, \citenamefont {Chou},\ and\ \citenamefont
		{Shih}}]{eom_persistent_2006}%
	\BibitemOpen
	\bibfield  {author} {\bibinfo {author} {\bibfnamefont {D.}~\bibnamefont
			{Eom}}, \bibinfo {author} {\bibfnamefont {S.}~\bibnamefont {Qin}}, \bibinfo
		{author} {\bibfnamefont {M.-Y.}\ \bibnamefont {Chou}}, \ and\ \bibinfo
		{author} {\bibfnamefont {C.~K.}\ \bibnamefont {Shih}},\ }\bibfield  {title}
	{\enquote {\bibinfo {title} {Persistent superconductivity in ultrathin {Pb}
				films: A scanning tunneling spectroscopy study},}\ }\href {\doibase
		10.1103/PhysRevLett.96.027005} {\bibfield  {journal} {\bibinfo  {journal}
			{Phys. Rev. Lett.}\ }\textbf {\bibinfo {volume} {96}},\ \bibinfo {pages}
		{027005} (\bibinfo {year} {2006})}\BibitemShut {NoStop}%
	\bibitem [{\citenamefont {Daghero}\ \emph {et~al.}(2012)\citenamefont
		{Daghero}, \citenamefont {Paolucci}, \citenamefont {Sola}, \citenamefont
		{Tortello}, \citenamefont {Ummarino}, \citenamefont {Agosto}, \citenamefont
		{Gonnelli}, \citenamefont {Nair},\ and\ \citenamefont
		{Gerbaldi}}]{daghero_large_2012}%
	\BibitemOpen
	\bibfield  {author} {\bibinfo {author} {\bibfnamefont {D.}~\bibnamefont
			{Daghero}}, \bibinfo {author} {\bibfnamefont {F.}~\bibnamefont {Paolucci}},
		\bibinfo {author} {\bibfnamefont {A.}~\bibnamefont {Sola}}, \bibinfo {author}
		{\bibfnamefont {M.}~\bibnamefont {Tortello}}, \bibinfo {author}
		{\bibfnamefont {G.~A.}\ \bibnamefont {Ummarino}}, \bibinfo {author}
		{\bibfnamefont {M.}~\bibnamefont {Agosto}}, \bibinfo {author} {\bibfnamefont
			{R.~S.}\ \bibnamefont {Gonnelli}}, \bibinfo {author} {\bibfnamefont {J.~R.}\
			\bibnamefont {Nair}}, \ and\ \bibinfo {author} {\bibfnamefont
			{C.}~\bibnamefont {Gerbaldi}},\ }\bibfield  {title} {\enquote {\bibinfo
			{title} {Large {{Conductance Modulation}} of {{Gold Thin Films}} by {{Huge
						Charge Injection}} via {{Electrochemical Gating}}},}\ }\href {\doibase
		10.1103/PhysRevLett.108.066807} {\bibfield  {journal} {\bibinfo  {journal}
			{Phys. Rev. Lett.}\ }\textbf {\bibinfo {volume} {108}},\ \bibinfo {pages}
		{066807} (\bibinfo {year} {2012})}\BibitemShut {NoStop}%
	\bibitem [{\citenamefont {Jeong}\ \emph {et~al.}(2013)\citenamefont {Jeong},
		\citenamefont {Aetukuri}, \citenamefont {Graf}, \citenamefont {Schladt},
		\citenamefont {Samant},\ and\ \citenamefont
		{Parkin}}]{jeong_suppression_2013}%
	\BibitemOpen
	\bibfield  {author} {\bibinfo {author} {\bibfnamefont {J.}~\bibnamefont
			{Jeong}}, \bibinfo {author} {\bibfnamefont {N.}~\bibnamefont {Aetukuri}},
		\bibinfo {author} {\bibfnamefont {T.}~\bibnamefont {Graf}}, \bibinfo {author}
		{\bibfnamefont {T.~D.}\ \bibnamefont {Schladt}}, \bibinfo {author}
		{\bibfnamefont {M.~G.}\ \bibnamefont {Samant}}, \ and\ \bibinfo {author}
		{\bibfnamefont {S.~S.~P.}\ \bibnamefont {Parkin}},\ }\bibfield  {title}
	{\enquote {\bibinfo {title} {Suppression of metal-insulator transition in
				{VO$_2$} by electric field-induced oxygen vacancy formation},}\ }\href
	{\doibase 10.1126/science.1230512} {\bibfield  {journal} {\bibinfo  {journal}
			{Science}\ }\textbf {\bibinfo {volume} {339}},\ \bibinfo {pages} {1402--1405}
		(\bibinfo {year} {2013})}\BibitemShut {NoStop}%
	\bibitem [{\citenamefont {Lu}\ \emph {et~al.}(2017)\citenamefont {Lu},
		\citenamefont {Zheliuk}, \citenamefont {Chen}, \citenamefont {Leermakers},
		\citenamefont {Hussey}, \citenamefont {Zeitler},\ and\ \citenamefont
		{Ye}}]{lu_full_2017}%
	\BibitemOpen
	\bibfield  {author} {\bibinfo {author} {\bibfnamefont {J.~M.}\ \bibnamefont
			{Lu}}, \bibinfo {author} {\bibfnamefont {O.}~\bibnamefont {Zheliuk}},
		\bibinfo {author} {\bibfnamefont {Q.~H.}\ \bibnamefont {Chen}}, \bibinfo
		{author} {\bibfnamefont {I.}~\bibnamefont {Leermakers}}, \bibinfo {author}
		{\bibfnamefont {N.~E.}\ \bibnamefont {Hussey}}, \bibinfo {author}
		{\bibfnamefont {U.}~\bibnamefont {Zeitler}}, \ and\ \bibinfo {author}
		{\bibfnamefont {J.~T.}\ \bibnamefont {Ye}},\ }\bibfield  {title} {\enquote
		{\bibinfo {title} {A full superconducting dome of strong {{Ising}} protection
				in gated monolayer {{WS$_2$}}},}\ }\href@noop {} {\bibfield  {journal}
		{\bibinfo  {journal} {arXiv:1703.06369 [cond-mat]}\ } (\bibinfo {year}
		{2017})},\ \Eprint {http://arxiv.org/abs/1703.06369} {arXiv:1703.06369
		[cond-mat]} \BibitemShut {NoStop}%
	\bibitem [{\citenamefont {Hayashi}\ \emph {et~al.}(1998)\citenamefont
		{Hayashi}, \citenamefont {Isoshima}, \citenamefont {Ichioka},\ and\
		\citenamefont {Machida}}]{hayashi_low-lying_1998}%
	\BibitemOpen
	\bibfield  {author} {\bibinfo {author} {\bibfnamefont {N.}~\bibnamefont
			{Hayashi}}, \bibinfo {author} {\bibfnamefont {T.}~\bibnamefont {Isoshima}},
		\bibinfo {author} {\bibfnamefont {M.}~\bibnamefont {Ichioka}}, \ and\
		\bibinfo {author} {\bibfnamefont {K.}~\bibnamefont {Machida}},\ }\bibfield
	{title} {\enquote {\bibinfo {title} {Low-{{Lying Quasiparticle Excitations}}
				around a {{Vortex Core}} in {{Quantum Limit}}},}\ }\href {\doibase
		10.1103/PhysRevLett.80.2921} {\bibfield  {journal} {\bibinfo  {journal}
			{Phys. Rev. Lett.}\ }\textbf {\bibinfo {volume} {80}},\ \bibinfo {pages}
		{2921--2924} (\bibinfo {year} {1998})}\BibitemShut {NoStop}%
	\bibitem [{\citenamefont {Dalla~Torre}\ \emph {et~al.}(2016)\citenamefont
		{Dalla~Torre}, \citenamefont {Benjamin}, \citenamefont {He}, \citenamefont
		{Dentelski},\ and\ \citenamefont {Demler}}]{dalla_torre_friedel_2016}%
	\BibitemOpen
	\bibfield  {author} {\bibinfo {author} {\bibfnamefont {E.~G.}\ \bibnamefont
			{Dalla~Torre}}, \bibinfo {author} {\bibfnamefont {D.}~\bibnamefont
			{Benjamin}}, \bibinfo {author} {\bibfnamefont {Y.}~\bibnamefont {He}},
		\bibinfo {author} {\bibfnamefont {D.}~\bibnamefont {Dentelski}}, \ and\
		\bibinfo {author} {\bibfnamefont {E.}~\bibnamefont {Demler}},\ }\bibfield
	{title} {\enquote {\bibinfo {title} {Friedel oscillations as a probe of
				fermionic quasiparticles},}\ }\href {\doibase 10.1103/PhysRevB.93.205117}
	{\bibfield  {journal} {\bibinfo  {journal} {Phys. Rev. B}\ }\textbf {\bibinfo
			{volume} {93}},\ \bibinfo {pages} {205117} (\bibinfo {year}
		{2016})}\BibitemShut {NoStop}%
	\bibitem [{\citenamefont {Machida}\ and\ \citenamefont
		{Koyama}(2003)}]{machida_friedel_2003}%
	\BibitemOpen
	\bibfield  {author} {\bibinfo {author} {\bibfnamefont {M.}~\bibnamefont
			{Machida}}\ and\ \bibinfo {author} {\bibfnamefont {T.}~\bibnamefont
			{Koyama}},\ }\bibfield  {title} {\enquote {\bibinfo {title} {Friedel
				{{Oscillation}} in {{Charge Profile}} and {{Position Dependent Screening}}
				around a {{Superconducting Vortex Core}}},}\ }\href {\doibase
		10.1103/PhysRevLett.90.077003} {\bibfield  {journal} {\bibinfo  {journal}
			{Phys. Rev. Lett.}\ }\textbf {\bibinfo {volume} {90}},\ \bibinfo {pages}
		{077003} (\bibinfo {year} {2003})}\BibitemShut {NoStop}%
	\bibitem [{\citenamefont {de~Gennes}(1964)}]{de_gennes_boundary_1964}%
	\BibitemOpen
	\bibfield  {author} {\bibinfo {author} {\bibfnamefont {P.~G.}\ \bibnamefont
			{de~Gennes}},\ }\bibfield  {title} {\enquote {\bibinfo {title} {Boundary
				effects in superconductors},}\ }\href {\doibase 10.1103/RevModPhys.36.225}
	{\bibfield  {journal} {\bibinfo  {journal} {Rev. Mod. Phys.}\ }\textbf
		{\bibinfo {volume} {36}},\ \bibinfo {pages} {225--237} (\bibinfo {year}
		{1964})}\BibitemShut {NoStop}%
	\bibitem [{\citenamefont {Covaci}\ and\ \citenamefont
		{Marsiglio}(2006)}]{covaci_proximity_2006}%
	\BibitemOpen
	\bibfield  {author} {\bibinfo {author} {\bibfnamefont {L.}~\bibnamefont
			{Covaci}}\ and\ \bibinfo {author} {\bibfnamefont {F.}~\bibnamefont
			{Marsiglio}},\ }\bibfield  {title} {\enquote {\bibinfo {title} {Proximity
				effect and {{Josephson}} current in clean strong/weak/strong superconducting
				trilayers},}\ }\href {\doibase 10.1103/PhysRevB.73.014503} {\bibfield
		{journal} {\bibinfo  {journal} {Phys. Rev. B}\ }\textbf {\bibinfo {volume}
			{73}},\ \bibinfo {pages} {014503} (\bibinfo {year} {2006})}\BibitemShut
	{NoStop}%
	\bibitem [{\citenamefont {Shanenko}\ \emph {et~al.}(2015)\citenamefont
		{Shanenko}, \citenamefont {Aguiar}, \citenamefont {Vagov}, \citenamefont
		{Croitoru},\ and\ \citenamefont {Milo\ifmmode \check{s}\else
			\v{s}\fi{}evi\ifmmode~\acute{c}\else \'{c}\fi{}}}]{shanenko_atomically_2015}%
	\BibitemOpen
	\bibfield  {author} {\bibinfo {author} {\bibfnamefont {A.~A.}\ \bibnamefont
			{Shanenko}}, \bibinfo {author} {\bibfnamefont {J.~A.}\ \bibnamefont
			{Aguiar}}, \bibinfo {author} {\bibfnamefont {A.}~\bibnamefont {Vagov}},
		\bibinfo {author} {\bibfnamefont {M.~D.}\ \bibnamefont {Croitoru}}, \ and\
		\bibinfo {author} {\bibfnamefont {M.~V.}\ \bibnamefont {Milo\ifmmode
				\check{s}\else \v{s}\fi{}evi\ifmmode~\acute{c}\else \'{c}\fi{}}},\ }\bibfield
	{title} {\enquote {\bibinfo {title} {Atomically flat superconducting
				nanofilms: Multiband properties and mean-field theory},}\ }\href {\doibase
		10.1088/0953-2048/28/5/054001} {\bibfield  {journal} {\bibinfo  {journal}
			{Supercond. Sci. Technol.}\ }\textbf {\bibinfo {volume} {28}},\ \bibinfo
		{pages} {054001} (\bibinfo {year} {2015})}\BibitemShut {NoStop}%
	\bibitem [{\citenamefont {Luh}\ \emph {et~al.}(2002)\citenamefont {Luh},
		\citenamefont {Miller}, \citenamefont {Paggel},\ and\ \citenamefont
		{Chiang}}]{luh_large_2002}%
	\BibitemOpen
	\bibfield  {author} {\bibinfo {author} {\bibfnamefont {D.-A.}\ \bibnamefont
			{Luh}}, \bibinfo {author} {\bibfnamefont {T.}~\bibnamefont {Miller}},
		\bibinfo {author} {\bibfnamefont {J.~J.}\ \bibnamefont {Paggel}}, \ and\
		\bibinfo {author} {\bibfnamefont {T.-C.}\ \bibnamefont {Chiang}},\ }\bibfield
	{title} {\enquote {\bibinfo {title} {Large electron-phonon coupling at an
				interface},}\ }\href {\doibase 10.1103/PhysRevLett.88.256802} {\bibfield
		{journal} {\bibinfo  {journal} {Phys. Rev. Lett.}\ }\textbf {\bibinfo
			{volume} {88}},\ \bibinfo {pages} {256802} (\bibinfo {year}
		{2002})}\BibitemShut {NoStop}%
	\bibitem [{\citenamefont {Chen}\ \emph
		{et~al.}(2012{\natexlab{b}})\citenamefont {Chen}, \citenamefont {Shanenko},\
		and\ \citenamefont {Peeters}}]{chen_superconducting_2012-1}%
	\BibitemOpen
	\bibfield  {author} {\bibinfo {author} {\bibfnamefont {Y.}~\bibnamefont
			{Chen}}, \bibinfo {author} {\bibfnamefont {A.~A.}\ \bibnamefont {Shanenko}},
		\ and\ \bibinfo {author} {\bibfnamefont {F.~M.}\ \bibnamefont {Peeters}},\
	}\bibfield  {title} {\enquote {\bibinfo {title} {Superconducting transition
				temperature of {Pb} nanofilms: Impact of thickness-dependent oscillations of
				the phonon-mediated electron-electron coupling},}\ }\href {\doibase
		10.1103/PhysRevB.85.224517} {\bibfield  {journal} {\bibinfo  {journal} {Phys.
				Rev. B}\ }\textbf {\bibinfo {volume} {85}},\ \bibinfo {pages} {224517}
		(\bibinfo {year} {2012}{\natexlab{b}})}\BibitemShut {NoStop}%
	\bibitem [{\citenamefont {Zhang}\ \emph {et~al.}(2005)\citenamefont {Zhang},
		\citenamefont {Jia}, \citenamefont {Han}, \citenamefont {Tang}, \citenamefont
		{Shen}, \citenamefont {Guo}, \citenamefont {Qiu},\ and\ \citenamefont
		{Xue}}]{zhang_band_2005-1}%
	\BibitemOpen
	\bibfield  {author} {\bibinfo {author} {\bibfnamefont {Y.-F.}\ \bibnamefont
			{Zhang}}, \bibinfo {author} {\bibfnamefont {J.-F.}\ \bibnamefont {Jia}},
		\bibinfo {author} {\bibfnamefont {T.-Z.}\ \bibnamefont {Han}}, \bibinfo
		{author} {\bibfnamefont {Z.}~\bibnamefont {Tang}}, \bibinfo {author}
		{\bibfnamefont {Q.-T.}\ \bibnamefont {Shen}}, \bibinfo {author}
		{\bibfnamefont {Y.}~\bibnamefont {Guo}}, \bibinfo {author} {\bibfnamefont
			{Z.~Q.}\ \bibnamefont {Qiu}}, \ and\ \bibinfo {author} {\bibfnamefont
			{Q.-K.}\ \bibnamefont {Xue}},\ }\bibfield  {title} {\enquote {\bibinfo
			{title} {Band structure and oscillatory electron-phonon coupling of {Pb} thin
				films determined by atomic-layer-resolved quantum-well states},}\ }\href
	{\doibase 10.1103/PhysRevLett.95.096802} {\bibfield  {journal} {\bibinfo
			{journal} {Phys. Rev. Lett.}\ }\textbf {\bibinfo {volume} {95}},\ \bibinfo
		{pages} {096802} (\bibinfo {year} {2005})}\BibitemShut {NoStop}%
	\bibitem [{\citenamefont {Blatt}\ and\ \citenamefont
		{Thompson}(1963)}]{blatt_shape_1963}%
	\BibitemOpen
	\bibfield  {author} {\bibinfo {author} {\bibfnamefont {J.~M.}\ \bibnamefont
			{Blatt}}\ and\ \bibinfo {author} {\bibfnamefont {C.~J.}\ \bibnamefont
			{Thompson}},\ }\bibfield  {title} {\enquote {\bibinfo {title} {Shape
				resonances in superconducting thin films},}\ }\href {\doibase
		10.1103/PhysRevLett.10.332} {\bibfield  {journal} {\bibinfo  {journal} {Phys.
				Rev. Lett.}\ }\textbf {\bibinfo {volume} {10}},\ \bibinfo {pages} {332--334}
		(\bibinfo {year} {1963})}\BibitemShut {NoStop}%
	\bibitem [{\citenamefont {Valentinis}\ \emph
		{et~al.}(2016{\natexlab{a}})\citenamefont {Valentinis}, \citenamefont
		{van~der Marel},\ and\ \citenamefont {Berthod}}]{valentinis_rise_2016}%
	\BibitemOpen
	\bibfield  {author} {\bibinfo {author} {\bibfnamefont {D.}~\bibnamefont
			{Valentinis}}, \bibinfo {author} {\bibfnamefont {D.}~\bibnamefont {van~der
				Marel}}, \ and\ \bibinfo {author} {\bibfnamefont {C.}~\bibnamefont
			{Berthod}},\ }\bibfield  {title} {\enquote {\bibinfo {title} {Rise and fall
				of shape resonances in thin films of {BCS} superconductors},}\ }\href
	{\doibase 10.1103/PhysRevB.94.054516} {\bibfield  {journal} {\bibinfo
			{journal} {Phys. Rev. B}\ }\textbf {\bibinfo {volume} {94}},\ \bibinfo
		{pages} {054516} (\bibinfo {year} {2016}{\natexlab{a}})}\BibitemShut
	{NoStop}%
	\bibitem [{\citenamefont {Valentinis}\ \emph
		{et~al.}(2016{\natexlab{b}})\citenamefont {Valentinis}, \citenamefont
		{van~der Marel},\ and\ \citenamefont {Berthod}}]{valentinis_bcs_2016}%
	\BibitemOpen
	\bibfield  {author} {\bibinfo {author} {\bibfnamefont {D.}~\bibnamefont
			{Valentinis}}, \bibinfo {author} {\bibfnamefont {D.}~\bibnamefont {van~der
				Marel}}, \ and\ \bibinfo {author} {\bibfnamefont {C.}~\bibnamefont
			{Berthod}},\ }\bibfield  {title} {\enquote {\bibinfo {title} {{BCS}
				superconductivity near the band edge: Exact results for one and several
				bands},}\ }\href {\doibase 10.1103/PhysRevB.94.024511} {\bibfield  {journal}
		{\bibinfo  {journal} {Phys. Rev. B}\ }\textbf {\bibinfo {volume} {94}},\
		\bibinfo {pages} {024511} (\bibinfo {year} {2016}{\natexlab{b}})}\BibitemShut
	{NoStop}%
	\bibitem [{\citenamefont {Hwang}\ \emph {et~al.}(2000)\citenamefont {Hwang},
		\citenamefont {Das~Sarma},\ and\ \citenamefont {Stroscio}}]{hwang_role_2000}%
	\BibitemOpen
	\bibfield  {author} {\bibinfo {author} {\bibfnamefont {E.~H.}\ \bibnamefont
			{Hwang}}, \bibinfo {author} {\bibfnamefont {S.}~\bibnamefont {Das~Sarma}}, \
		and\ \bibinfo {author} {\bibfnamefont {M.~A.}\ \bibnamefont {Stroscio}},\
	}\bibfield  {title} {\enquote {\bibinfo {title} {Role of confined phonons in
				thin-film superconductivity},}\ }\href {\doibase 10.1103/PhysRevB.61.8659}
	{\bibfield  {journal} {\bibinfo  {journal} {Phys. Rev. B}\ }\textbf {\bibinfo
			{volume} {61}},\ \bibinfo {pages} {8659--8662} (\bibinfo {year}
		{2000})}\BibitemShut {NoStop}%
	\bibitem [{\citenamefont {Saniz}\ \emph {et~al.}(2013)\citenamefont {Saniz},
		\citenamefont {Partoens},\ and\ \citenamefont
		{Peeters}}]{saniz_confinement_2013}%
	\BibitemOpen
	\bibfield  {author} {\bibinfo {author} {\bibfnamefont {R.}~\bibnamefont
			{Saniz}}, \bibinfo {author} {\bibfnamefont {B.}~\bibnamefont {Partoens}}, \
		and\ \bibinfo {author} {\bibfnamefont {F.~M.}\ \bibnamefont {Peeters}},\
	}\bibfield  {title} {\enquote {\bibinfo {title} {Confinement effects on
				electron and phonon degrees of freedom in nanofilm superconductors: A Green
				function approach},}\ }\href {\doibase 10.1103/PhysRevB.87.064510} {\bibfield
		{journal} {\bibinfo  {journal} {Phys. Rev. B}\ }\textbf {\bibinfo {volume}
			{87}},\ \bibinfo {pages} {064510} (\bibinfo {year} {2013})}\BibitemShut
	{NoStop}%
	\bibitem [{\citenamefont {Qi}\ \emph {et~al.}(2007)\citenamefont {Qi},
		\citenamefont {Ma}, \citenamefont {Jiang}, \citenamefont {Ji}, \citenamefont
		{Fu}, \citenamefont {Jia}, \citenamefont {Xue},\ and\ \citenamefont
		{Zhang}}]{qi_atomic-layer-resolved_2007}%
	\BibitemOpen
	\bibfield  {author} {\bibinfo {author} {\bibfnamefont {Y.}~\bibnamefont
			{Qi}}, \bibinfo {author} {\bibfnamefont {X.}~\bibnamefont {Ma}}, \bibinfo
		{author} {\bibfnamefont {P.}~\bibnamefont {Jiang}}, \bibinfo {author}
		{\bibfnamefont {S.}~\bibnamefont {Ji}}, \bibinfo {author} {\bibfnamefont
			{Y.}~\bibnamefont {Fu}}, \bibinfo {author} {\bibfnamefont {J.-F.}\
			\bibnamefont {Jia}}, \bibinfo {author} {\bibfnamefont {Q.-K.}\ \bibnamefont
			{Xue}}, \ and\ \bibinfo {author} {\bibfnamefont {S.~B.}\ \bibnamefont
			{Zhang}},\ }\bibfield  {title} {\enquote {\bibinfo {title}
			{Atomic-layer-resolved local work functions of {Pb} thin films and their
				dependence on quantum well states},}\ }\href {\doibase 10.1063/1.2403926}
	{\bibfield  {journal} {\bibinfo  {journal} {Appl. Phys. Lett.}\ }\textbf
		{\bibinfo {volume} {90}},\ \bibinfo {pages} {013109} (\bibinfo {year}
		{2007})}\BibitemShut {NoStop}%
	\bibitem [{\citenamefont {Yu}\ \emph {et~al.}(1976)\citenamefont {Yu},
		\citenamefont {Strongin},\ and\ \citenamefont {Paskin}}]{yu_consistent_1976}%
	\BibitemOpen
	\bibfield  {author} {\bibinfo {author} {\bibfnamefont {M.}~\bibnamefont
			{Yu}}, \bibinfo {author} {\bibfnamefont {M.}~\bibnamefont {Strongin}}, \ and\
		\bibinfo {author} {\bibfnamefont {A.}~\bibnamefont {Paskin}},\ }\bibfield
	{title} {\enquote {\bibinfo {title} {Consistent calculation of boundary
				effects in thin superconducting films},}\ }\href {\doibase
		10.1103/PhysRevB.14.996} {\bibfield  {journal} {\bibinfo  {journal} {Phys.
				Rev. B}\ }\textbf {\bibinfo {volume} {14}},\ \bibinfo {pages} {996--1001}
		(\bibinfo {year} {1976})}\BibitemShut {NoStop}%
	\bibitem [{\citenamefont {Roditchev}\ \emph {et~al.}(2015)\citenamefont
		{Roditchev}, \citenamefont {Brun}, \citenamefont {Serrier-Garcia},
		\citenamefont {Cuevas}, \citenamefont {Bessa}, \citenamefont {Milo{\v
				s}evi{\'c}}, \citenamefont {Debontridder}, \citenamefont {Stolyarov},\ and\
		\citenamefont {Cren}}]{roditchev_direct_2015}%
	\BibitemOpen
	\bibfield  {author} {\bibinfo {author} {\bibfnamefont {D.}~\bibnamefont
			{Roditchev}}, \bibinfo {author} {\bibfnamefont {C.}~\bibnamefont {Brun}},
		\bibinfo {author} {\bibfnamefont {L.}~\bibnamefont {Serrier-Garcia}},
		\bibinfo {author} {\bibfnamefont {J.~C.}\ \bibnamefont {Cuevas}}, \bibinfo
		{author} {\bibfnamefont {V.~H.~L.}\ \bibnamefont {Bessa}}, \bibinfo {author}
		{\bibfnamefont {M.~V.}\ \bibnamefont {Milo{\v s}evi{\'c}}}, \bibinfo {author}
		{\bibfnamefont {F.}~\bibnamefont {Debontridder}}, \bibinfo {author}
		{\bibfnamefont {V.}~\bibnamefont {Stolyarov}}, \ and\ \bibinfo {author}
		{\bibfnamefont {T.}~\bibnamefont {Cren}},\ }\bibfield  {title} {\enquote
		{\bibinfo {title} {Direct observation of {{Josephson}} vortex cores},}\
	}\href {\doibase 10.1038/nphys3240} {\bibfield  {journal} {\bibinfo
			{journal} {Nat. Phys.}\ }\textbf {\bibinfo {volume} {11}},\ \bibinfo {pages}
		{332--337} (\bibinfo {year} {2015})}\BibitemShut {NoStop}%
\end{thebibliography}
%

\end{document}